\documentstyle[eqsecnum,aps]{revtex}

\begin{document}

\thispagestyle{empty}
{\baselineskip-4pt
\font\yitp=cmmib10 scaled\magstep2
\font\elevenmib=cmmib10 scaled\magstep1  \skewchar\elevenmib='177
\leftline{\baselineskip20pt
\vbox to0pt
   { {\yitp\hbox{Osaka \hspace{1.5mm} University} }
     {\large\sl\hbox{{Theoretical Astrophysics}} }\vss}}

\rightline{\large\baselineskip20pt\rm\vbox to20pt{
\baselineskip14pt
\hbox{OU-TAP 38}
\hbox{KUNS 1394}
\hbox{gr-qc/??}
\vspace{2mm}
\hbox{June, 1996}\vss}}

\vspace{2cm} 

\begin{center}
{\Large\bf Gravitational Radiation Reaction} 
{\Large\bf to a Particle Motion}
\end{center}
\bigskip

\centerline{\large Yasushi Mino,$^{1,2}$\footnote{Electronic address: 
mino@vega.ess.sci.osaka-u.ac.jp} $~$Misao Sasaki,$^1$\footnote{
Electronic address: misao@vega.ess.sci.osaka-u.ac.jp} 
and Takahiro Tanaka$^1$\footnote{Electronic address: 
tama@vega.ess.sci.osaka-u.ac.jp}}
\bigskip
\begin{center}{\em $^1$Department of Earth and Space Science, 
Graduate School of Science} \\
{\em  Osaka University, Toyonaka 560, Japan}\\
{\em $^2$Department of Physics, Faculty of Science, 
Kyoto University, Kyoto 606-01, Japan}
\end{center}

\bigskip

\begin{abstract}
A small mass particle traveling in a curved spacetime 
is known to trace a background geodesic in the 
lowest order approximation with respect to the 
particle mass. In this paper, we discuss the leading 
order correction to the equation of motion of the 
particle, which presumably describes the effect of
gravitational radiation reaction.

We derive the equation of motion in two different ways. 
The first one is an extension of the well-known 
formalism by DeWitt and Brehme developed for deriving
 the equation of motion of an electrically charged 
particle. Constructing the conserved rank two symmetric tensor, 
and integrating it over the interior of 
the world tube surrounding the orbit, we derive 
the equation of motion. Although the 
calculation in this approach is straightforward, it 
contains less rigorous points. In contrast to the 
electromagnetic case, in which there are two different 
charges, i.e., the electric charge and the mass, 
the gravitational counterpart has only one charge.
This fact prevents us from using the same renormalization 
scheme that was used in the electromagnetic case. In order 
to overcome this difficulty, we put an ansatz in evaluating 
the integral of the conserved tensor on a three spatial volume 
which defines the momentum of the small particle. 

To make clear the subtlety in the first approach, we then consider 
the asymptotic matching of two different schemes, i.e., the 
internal scheme in which the small particle is represented by 
a spherically symmetric black hole with tidal perturbations and 
the external scheme 
in which the metric is given by small perturbations on the 
given background geometry. 
The equation of motion is obtained from the consistency 
condition of the matching. 
We find that in both ways the same equation of motion is obtained. 
The resulting equation of motion is analogous to that derived 
in the electromagnetic case. 
We discuss implications of this equation of motion.

\noindent
PACS number(s): 04.30.Db, 04.25.-g
\end{abstract}


\section{Introduction} 
The gravitational waves from an inspiralling binary is 
one of the most promising sources expected to be detected 
by the near-future interferometric gravitational wave 
detectors such as LIGO, VIRGO and LISA\cite{Thorne,LISA}.  
In order to extract the information of the binaries from 
the last inspiralling stage, 
we need accurate theoretical templates of the gravitational 
wave forms\cite{Last}. 
As an approach, perturbations of a black hole by an orbiting 
small particle has been studied\cite{bhp}. 
In all the previous works, the rate of change of orbital parameters 
are assumed to be determined by the energy balance. 
Although this prescription is powerful, the background black hole 
is required to have a sufficient number of Killing vector fields 
in order to relate the outgoing gravitational waves 
with the rate of change of orbital parameters of the particle. 
The Kerr black hole is an important example which does not have 
a sufficient number of Killing vector fields and so 
rigorous discussions are restricted to the case in which 
the particle is in a circular or a equatorial orbit. 
Thus the effect of the radiation reaction 
on the motion of a small mass particle in 
a general curved spacetime is an important target of 
theoretical investigation. 

On the other hand, the effect of the electromagnetic radiation 
reaction on the motion of a charged particle in a curved spacetime 
was discussed by DeWitt and Brehme\cite{DeWitt} (Hereafter DB).
In the electromagnetic case, the total energy momentum tensor 
composed of the particle and field contributions satisfies the conservation 
law. Its divergence is integrated over the interior of the tube 
surrounding the particle orbit with infinitesimal length. 
The part of the integration which does not vanish in the 
small tube radius limit is transformed into 
the surface integrations over the both ends of the tube 
and over the surface of the tube by using the Gauss' theorem. 
The integrations over the top and bottom of the tube 
give the definition of the particle momenta at both ends and 
the difference between them 
represents the change of the momentum 
during this infinitesimal time interval, which is to be equated 
with the momentum flow given by the 
integration over the surface of the tube. 
In this way the equation of motion is obtained. 
As shown below, also in the case of gravitational radiation reactions, 
we can  construct a conserved rank two tensor 
defined on the background spacetime, 
composed of the matter field and the metric perturbation. 
However, there is an important difference between electromagnetic 
and gravitational cases. 
In the electromagnetism, we can consider an extended 
charge distribution which is supported by a certain force other 
than the electromagnetic field. 
Thus it will be natural to assume that the charge and mass 
distributions of a point-like particle are not distorted by 
the effect of the radiation reaction. 
Therefore one may consistently assume that 
the momentum and the electric current of the particle are 
proportional to the 4-velocity of the particle. 
Moreover the electromagnetic charge $e$ is not directly related 
to the energy momentum of the particle which is proportional 
to the mass $m$. 
Hence, even if the limit of zero particle radius 
is taken, the divergent self-energy, $(\propto e^2)$ 
can be renormalized by shifting the bare mass. 
In the case of the gravitational radiation reaction, 
it is not possible to consider such an ideal point-like 
particle because every force field universally couples with gravity. 
Even worse, the role of $e$ in the electromagnetism 
is also attributed to $m$. 
Thus the simple renormalization 
scheme does not make any sense. 
In order to overcome this difficulty, we put an ansatz 
that the particle momentum, defined in a similar way
as in the electromagnetism but without taking the 
limit of the small particle radius, be proportional 
to the 4-velocity of the particle, 
which will not be justified within this framework. 
Under this assumption, we obtain an equation 
of motion in the covariant form with respect 
to the background spacetime which is analogous 
to that obtained in the case of electromagnetic case. 

In order to develop a more rigorous formalism, 
we consider an extension of the matched asymptotic
expansion that has been already developed by many authors (e.g.,
D'Eath \cite{Death} and Thorne and Hartle \cite{Thorne1}). 
We assume that the internal metric which describes the 
geometry around the particle is represented by 
a Schwarzschild black hole of small mass in the lowest 
order approximation. 
As the particle moves in the curved background, it 
suffers from the tidal distortion. 
This effect is taken into account by the 
homogeneous linear perturbations of the black hole. 
Since we know $\ell=0$ and $1$ homogeneous perturbations of the black hole 
are purely indebted to gauge degrees of freedom as long 
as both the mass and angular momentum of the black hole stay constant, 
we set them to vanish. 
We also assume that the external metric is approximated by 
the linear perturbations on the background spacetime 
generated by a point-like particle. 
Then we consider a limited class of coordinate 
transformation which keeps the meaning 
of the center of the particle unambiguous 
and match the external metric with the internal one 
in the matching region where the both 
approximations are valid. 
Then we find that for an arbitrary orbit 
the consistent coordinate transformation does not exist, 
and this consistency condition determines 
the equation of motion, which is no different from 
that obtained in the first approach. 

This paper is organized as follows. 
In section 2  we explain the covariant expansion 
method of the tensor Green function, 
which becomes important in both approaches discussed 
in the succeeding two sections. 
In section 3 we discuss the first approach, i.e., 
an extension of DeWitt and Brehme's 
electromagnetic radiation reaction equation of motion 
to the gravitational counter part. 
Section 4 is devoted to the second approach, i.e.,  
the matched asymptotic expansion method to derive 
the equation of motion of a small black hole. 
We find both approaches give the same equation of motion.
In section 5 we discuss implications of the result.
Section 6 is devoted to conclusion.

We suppose the reader is familiar with the concept 
of `bi-tensors' and some useful tools developed by DB. 
Following DB, we assign the indices $\alpha,\beta,\gamma,\delta, 
\epsilon,\zeta,\eta$ for the point on the particle
trajectory, $z(\tau)$, and the indices $\mu,\nu,\xi,\rho,\sigma$ 
for the field point, $x$.
For the reader's convenience, the notation and basic formulas are
summarized in Appendix A.


\section{Metric Perturbation}

In this section, for the later use, 
we calculate the metric perturbation, $\delta g_{\mu\nu}$, 
induced by a point-like particle on the background metric, $g_{\mu\nu}$.
The background metric is assumed to satisfy 
the vacuum Einstein equations. 
Thus, in the following calculations, we use the fact that 
the background Ricci tensor vanishes. 
As we assume that the particle mass, $m$, is small 
compared with the background curvature scale, $L$,
we approximate $\delta g_{\mu\nu}$ 
by the linear perturbation induced by a point-like particle, 
$h_{\mu\nu}$, in the whole spacetime region 
except for the vicinity of the world line of the particle. 
We call the region in which this approximation 
is valid as the external zone.  
On the other hand, 
in the vicinity of the world line of the particle, 
the self gravity of the particle dominates and 
the metric cannot be described by the linear perturbation 
induced by a point-like particle. 
We call this region as the internal zone. 
In this section we concentrate on the external zone. 
The calculation is performed in an analogous manner 
to the case of the scalar and vector perturbations 
developed by DB.

Here we consider the linearized Einstein equations. 
We introduce the trace-reversed metric perturbation, 
\begin{eqnarray}
\psi_{\mu\nu}(x) &=& h_{\mu\nu}(x) -{1\over 2}g_{\mu\nu}(x) h(x), 
\end{eqnarray}
and set the harmonic gauge condition,
\begin{eqnarray}
\psi^{\mu\nu}{}_{;\nu}(x)&=&0,
\end{eqnarray}
where $h(x)$ and $\psi(x)$ are 
the trace of $h^{\mu\nu}(x)$ and that of $\psi^{\mu\nu}(x)$, respectively, 
and the semicolon means the covariant derivative with respect to the 
background metric. In this gauge, the linearized Einstein equations become
\begin{eqnarray}
-{1\over 2}\psi^{\mu\nu;\xi}{}_\xi(x)
-R^\mu{}_\xi{}^\nu{}_\rho (x) \psi^{\xi\rho}(x)
= 8 \pi G T^{\mu\nu}(x) +O({\bf h}^2),
\end{eqnarray}
where $O({\bf h}^2)$ stands for terms quadratic or of higher powers
in the metric perturbation.
Thus we define the tensor Green function $G^{\mu\nu\alpha\beta}(x,z)$ 
which satisfies 
\begin{eqnarray}
G^{\mu\nu\alpha\beta;\xi}{}_{;\xi}(x,z)
+2R^\mu{}_\xi{}^\nu{}_\rho(x) G^{\xi\rho\alpha\beta}(x,z) 
=-2\bar g^{\alpha(\mu}(x,z)\bar g^{\nu)\beta}(x,z)
{\delta^{(4)}(z-x) \over \sqrt{-g}}, 
\label{eq:green}
\end{eqnarray}
where $\bar g^{\mu\alpha}(x,z)$ is the bi-vector of geodetic 
parallel displacement defined in Eq.~(\ref{131db}) of Appendix A and 
$g$ is the determinant of the metric $g_{\mu\nu}(x)$. 

First we consider 
the elementary solution $G_{*}^{\mu\nu\alpha\beta}(x,z)$ which satisfies 
Eq.~(\ref{eq:green}) except at the $\sigma(x,z)\rightarrow 0$ limit and 
takes the Hadamard form, 
\begin{eqnarray}
G_*^{\mu\nu\alpha\beta}(x,z)=
{1\over (2\pi)^2}\left({u^{\mu\nu\alpha\beta}(x,z)\over\sigma(x,z)}
+v^{\mu\nu\alpha\beta}(x,z)\log|\sigma(x,z)|
+w^{\mu\nu\alpha\beta}(x,z)\right), 
\label{eq:Hadamard}
\end{eqnarray}
where $\sigma(x,z)$ is a half of the square of geodetic interval
which is defined in Eq.~(1.10) of DB. Its property is summarized in
Eq.~(\ref{111db}) of Appendix A.
The functions $u^{\mu\nu\alpha\beta}(x,z)$, $v^{\mu\nu\alpha\beta}(x,z)$ and 
$w^{\mu\nu\alpha\beta}(x,z)$ are bi-tensors 
which are regular in the $\sigma(x,z)\rightarrow 0$ limit and  
$u^{\mu\nu\alpha\beta}(x,z)$ satisfies the normalization 
condition, 
\begin{eqnarray}
\lim_{x\rightarrow z}u^{\mu\nu\alpha\beta}(x,z)= 
\lim_{x\rightarrow z}2\bar g^{\alpha(\mu}(x,z)\bar g^{\nu)\beta}(x,z). 
\label{eq:unorm} 
\end{eqnarray}
If we put the form (\ref{eq:Hadamard}) into the left hand side of 
Eq.~(\ref{eq:green}), the terms can be classified into three parts.  
One is the terms which contain the factor $1/\sigma^2(x,z)$ manifestly 
and another is the terms which contain $\log|\sigma(x,z)|$. 
The remaining terms have no singular behavior at the
$\sigma(x,z)\rightarrow 0$ limit. 
Since the form (\ref{eq:Hadamard}) is redundant, 
we can set these three sets to vanish separately:
\begin{eqnarray} 
&& \left(2u^{\mu\nu\alpha\beta;\xi}(x,z) 
-{\Delta^{;\xi}(x,z)\over\Delta(x,z)}u^{\mu\nu\alpha\beta}(x,z)\right) 
\sigma_{;\xi}(x,z) = 0, 
\label{eq:ueq}
\\ 
&& v^{\mu\nu\alpha\beta;\xi}{}_{;\xi}(x,z) 
+2R^\mu{}_\xi{}^\nu{}_\rho(x) v^{\xi\rho\alpha\beta}(x,z) = 0, 
\label{eq:veq}
\\ 
&& 2v^{\mu\nu\alpha\beta}(x,z) 
+\left(2v^{\mu\nu\alpha\beta;\xi}(x,z) 
-{\Delta^{;\xi}(x,z)\over\Delta(x,z)}v^{\mu\nu\alpha\beta}(x,z)\right) 
\sigma_{;\xi}(x,z) 
+u^{\mu\nu\alpha\beta;\xi}{}_{;\xi}(x,z) 
\nonumber \\ 
&& \qquad \quad +2R^\mu{}_\xi{}^\nu{}_\rho(x) u^{\xi\rho\alpha\beta}(x,z) 
+\left(w^{\mu\nu\alpha\beta;\xi}{}_{;\xi}(x,z)
+2R^\mu{}_\xi{}^\nu{}_\rho(x) w^{\xi\rho\alpha\beta}(x,z)\right)\sigma(x,z)
 = 0, 
\label{eq:weq} 
\end{eqnarray}
where we used 
the bi-scalar $\Delta(x,z)$ defined in Eq.~(\ref{Deltadef}) in 
Appendix A.
Equation~(\ref{eq:ueq}) is solved   
with the normalization (\ref{eq:unorm}) as 
\begin{eqnarray} 
u^{\mu\nu\alpha\beta}(x,z)=
2\bar g^{\alpha(\mu}(x,z)\bar g^{\nu)\beta}(x,z)\sqrt{\Delta(x,z)}.
\label{eq:u}
\end{eqnarray}
The functions $v^{\mu\nu\alpha\beta}(x,z)$ and $w^{\mu\nu\alpha\beta}(x,z)$ 
are to be determined by solving Eqs.~(\ref{eq:veq}) and (\ref{eq:weq}). 
The function $w^{\mu\nu\alpha\beta}(x,z)$ is not needed but 
the function $v^{\mu\nu\alpha\beta}(x,z)$ plays an important role in 
the following discussion. 
Although it is difficult to find 
the solution of $v^{\mu\nu\alpha\beta}(x,z)$ in an arbitrary 
background spacetime, 
its explicit form is not required for the succeeding discussions. 
However it is important to note that $v^{\mu\nu\alpha\beta}(x,z)$ is uniquely
determined. The reason is as follows.
{}From Eq.~(\ref{eq:veq}) one finds it satisfies a hyperbolic 
equation. Hence the problem is if its Cauchy data are unique or not.
First we note the coincidence limit of Eq.~(\ref{eq:weq}), which gives
\begin{eqnarray}
\lim_{x\rightarrow z}v^{\mu\nu\alpha\beta}(x,z)
=-\lim_{x\rightarrow z}2\bar g^\alpha{}_{(\xi} (z,x)
\bar g^{\beta}{}_{\rho)}(z,x)
R^{\mu\xi\nu\rho}(x). 
\label{eq:ulim}
\end{eqnarray}
Then taking the null limit $\sigma(x,z)\rightarrow 0$ of
 Eq.~(\ref{eq:weq}), 
we obtain the first order differential equation of 
$v^{\mu\nu\alpha\beta}(x,z)$ which can be solved along the 
null geodesic. 
Thus this equation with the boundary condition (\ref{eq:ulim}) 
uniquely determines $v^{\mu\nu\alpha\beta}(x,z)$ on the light cone 
emanating from $z$. 
Therefore the hyperbolic equation (\ref{eq:veq}) has a unique
solution. We also mention that $v^{\mu\nu\alpha\beta}(x,z)$ is divergence free,
\begin{equation}
v^{\mu\nu\alpha\beta}{}_{;\nu}(x,z)=0. 
\label{divv}
\end{equation}
To see this we note the harmonic gauge condition on the Green 
function requires 
\begin{equation}
 \lim_{\sigma\rightarrow 0} v^{\mu\nu\alpha\beta}{}_{;\nu}(x,z)=0.
\end{equation}
We also see that the equation for  
$v^{\mu\nu\alpha\beta}{}_{;\nu}(x,z)$ follows from Eq.~(\ref{eq:veq}),
\begin{equation}
 \left[v^{\mu\nu\alpha\beta}{}_{;\nu}(x,z)\right]{}^{;\xi}{}_{;\xi}=0, 
\end{equation}
where we have used the fact $R^{\mu\xi\nu\rho}{}_{;\rho}=0$, 
which is proved by contracting the Bianchi identities 
for the vacuum case. Thus we conclude
that Eq.~(\ref{divv}) holds everywhere.

The Feynman propagator $G_F^{\mu\nu\alpha\beta}(x,z)$ can be derived 
from the elementary solution $G_*^{\mu\nu\alpha\beta}(x,z)$ 
by the $i\epsilon$-prescription. 
\begin{eqnarray}
G_F^{\mu\nu\alpha\beta}(x,z)
={1\over (2\pi)^2}\left({u^{\mu\nu\alpha\beta}(x,z)\over\sigma(x,z)+i\epsilon}
+v^{\mu\nu\alpha\beta}(x,z)\log(\sigma(x,z)+i\epsilon)
+w^{\mu\nu\alpha\beta}(x,z)\right). 
\end{eqnarray}
The imaginary part of the Feynman propagator $G_F^{\mu\nu\alpha\beta}(x,z)$ 
gives the symmetric Green function $\bar G^{\mu\nu\alpha\beta}(x,z)$, 
from which we can obtain 
the retarded Green function $G_{Ret}^{\mu\nu\alpha\beta}(x,z)$, 
and the advanced Green function $G_{Adv}^{\mu\nu\alpha\beta}(x,z)$ as
\begin{eqnarray}
\bar G^{\mu\nu\alpha\beta}(x,z) 
&=&-{1\over 2}{\rm Im}\left[G_F^{\mu\nu\alpha\beta}(x,z)\right] \nonumber \\
&=&{1\over 8\pi}\left[u^{\mu\nu\alpha\beta}(x,z)\delta(\sigma(x,z))
-v^{\mu\nu\alpha\beta}(x,z)\theta(-\sigma(x,z))\right], \\
G_{Ret}^{\mu\nu\alpha\beta}(x,z)
&=&2\theta[\Sigma(x),z]\bar G^{\mu\nu\alpha\beta}(x,z), \\
G_{Adv}^{\mu\nu\alpha\beta}(x,z)
&=&2\theta[z,\Sigma(x)]\bar G^{\mu\nu\alpha\beta}(x,z), 
\end{eqnarray}
where 
$\Sigma(x)$ is an arbitrary space-like hypersurface 
containing $x$, and $\theta[\Sigma(x),z]=1
-\theta[z,\Sigma(x)]$ is equal to $1$ when $z$ lies in the past 
of $\Sigma(x)$ and vanishes when $z$ lies in the future.

Now, using the above obtained Green functions, 
we compute the trace-reversed metric perturbation $\psi^{\mu\nu}(x)$ 
induced by a point-like particle whose energy momentum tensor 
is given by 
\begin{eqnarray}
T^{\mu\nu}(x)=m\int d\tau~ v^\mu(x,\tau) v^\nu(x,\tau) 
{\delta^{(4)}(x-z(\tau))\over \sqrt{-g}}, 
\label{eq:point}
\end{eqnarray}
where $v^\mu(x,\tau)=\bar g^\mu{}_\alpha(x,z(\tau))\dot z^\alpha(\tau)$ 
and $\dot z^\alpha(\tau)=d z^{\alpha}/d\tau$. 
It is also assumed that the particle mass $m$ is small 
compared with the background curvature scale $L$. 
Throughout this paper, we take the unit in which 
$L$ becomes of $O(1)$.
There appear relations whose dimensionality looks 
wrong but in those cases $L$ is omitted for 
notational simplicity. 

At this point, we must comment on the reason why 
we consider the point-like particle. 
Even in the linear perturbation, in order 
to generate a general gravitational field 
in the external zone,  we need to consider 
a source with arbitrary higher multipole moments. 
However, in the following discussion, we 
are going to discuss the situation in which 
those higher moments are negligible.
Thus we consider this special case here. 
Later it will become clarified in what situation 
this assumption becomes consistent. 

Now, for the point-like particle, 
the trace-reversed metric perturbation $\psi^{\mu\nu}(x)$ 
with the retarded or advanced boundary condition is 
computed as 
\begin{eqnarray} 
\psi_{Ret/Adv}^{\mu\nu}(x) 
&=&\pm 2Gm\Biggl(\biggl[ 
{1\over\dot\sigma(x,z(\tau))}u^{\mu\nu}{}_{\alpha\beta}(x,z(\tau))
\dot z^\alpha(\tau) \dot z^\beta(\tau)\biggr]_{\tau=\tau_{Ret/Adv}(x)} 
\nonumber \\ && \qquad \qquad 
-\int^{\tau_{Ret/Adv}(x)}_{\mp\infty}
d\tau v^{\mu\nu}{}_{\alpha\beta}(x,z(\tau))
\dot z^\alpha(\tau)\dot z^\beta(\tau)\Biggr). 
\label{eq:metper} 
\end{eqnarray} 
where $\tau_{Ret/Adv}(x)$ is the retarded or advanced time 
of the particle and is a scalar function which is determined by 
\begin{eqnarray}
&& \sigma\left(x,z(\tau_{Ret/Adv})\right)=0, \\ 
&& \theta\left(\Sigma(x),z(\tau_{Ret})\right)
=\theta\left(z(\tau_{Adv}),\Sigma(x)\right)=1.
\end{eqnarray}
Throughout this paper, we take the convention that 
the upper sign is assigned for the 
retarded boundary condition
and the lower sign is for the advanced one. 

Since the expression (\ref{eq:metper}) 
containing the retarded or advanced time $\tau_{Ret/Adv}(x)$ 
is not convenient for the computations in the succeeding sections, 
we rewrite it by introducing a new parameterization of the 
field point $x$.
We foliate the spacetime with spacelike 3-surfaces 
perpendicular to the particle trajectory. 
More strictly, the 3-surfaces are defined as a one-parameter 
family of $\tau$ by the relation, 
$\sigma_{;\alpha}(x,z(\tau))\dot z^\alpha(\tau)=0$. 
We denote the value of $\tau$ of the 3-surface containing 
the point $x$ by $\tau_x$. 
That is
\begin{eqnarray}
\left[\sigma_{;\alpha}(x,z(\tau))\dot z^\alpha(\tau)\right]_{\tau=\tau_x}=0. 
\label{eq:foli}
\end{eqnarray}
To distinguish the spatial points on the same 3-surface, 
we use $\sigma_{;\alpha}(x,z(\tau_{x}))$ and  
denote the distance between $x$ and $z(\tau_x)$ 
by 
\begin{equation}
\epsilon(x):=\sqrt{2\sigma(x,z(\tau_x))}.
\end{equation} 
As we are interested in the region where the 
linear perturbation is valid, 
$\epsilon(x)$ must be much greater than $Gm$. 
However, concentrating on the region close to the 
particle even in the external zone, 
we can take $\epsilon(x)$ is smaller than the background curvature scale $L$,
because $Gm$ is assumed to be much smaller than $L$. 
Thus we can consider the expansion assuming that $\epsilon(x)$ is small.

In the following calculation there appear the higher derivatives of 
$\dot z$, such as $\ddot z$ and $\stackrel{...}{z}$, where 
a dot means the covariant derivative along the trajectory of 
the particle. 
Since we are considering the case in which  
the deviation of particle trajectory from the geodesic 
is a small correction due to the radiation reaction, 
we suppose that those higher derivatives are 
at most of $O(\epsilon(x))$. 
Thus we define the inverse of the reaction time scale 
$\tau_r^{-1}(<\epsilon(x))$ 
as the scale of $\ddot z$, $\stackrel{...}{z}$ and so on.  
We will see later that $\tau_r^{-1}\approx Gm$ and 
this assumption is found to be consistent. 

We first consider 
the time retardation or advancement, $\delta_{Ret/Adv}(x)$, 
\begin{eqnarray}
\delta_{Ret/Adv}(x):=\tau_{Ret/Adv}(x)-\tau_x. 
\end{eqnarray}
The time retardation or advancement, $\delta_{Ret/Adv}(x)$, is expanded 
with respect to $\epsilon(x)$, 
\begin{eqnarray}
\delta_{Ret/Adv}(x)&=&
\mp\epsilon(x)\kappa^{-1}(x)\biggl(1
\mp{1\over 6}\epsilon(x)\kappa^{-3}(x)
\stackrel{...}{z}{}^{\alpha}(\tau_x)\sigma_{;\alpha}(x,z(\tau_x))
\nonumber \\ && \qquad \qquad
-{1\over 24}\epsilon^2(x)\kappa^{-4}(x)\ddot z^2(\tau_x)\biggr)
+O(\epsilon^4), 
\label{eq:retard}
\end{eqnarray} 
as given in Eq.~(4.40) of DB
\footnote{In DB, $\sigma_{;\alpha}(x,z(\tau_x))$ is 
replaced by $-\epsilon(x)n_{i\alpha}(\tau_x)\Omega_i$.},
where $\kappa^2(x)=-[\ddot\sigma(x,z(\tau))]_{\tau=\tau_x}$.
Then the expression for the trace-reversed metric perturbation 
(\ref{eq:metper}) becomes
\begin{eqnarray}
\psi_{\mu\nu}(x)
&=&\pm 2Gm \bar g_{\mu\alpha}\bar g_{\nu\beta}
\nonumber \\ && \quad 
\Biggl(\pm{2\over \epsilon}\kappa^{-1}\dot z^\alpha\dot z^\beta
-4\dot z^{(\alpha}\ddot z^{\beta)}
-2\dot z^\gamma\sigma^{;\delta}\dot z^\epsilon 
R_{\gamma\delta\epsilon}{}^{(\alpha}\dot z{}^{\beta)} 
\mp 2\epsilon R^\alpha{}_\gamma{}^\beta{}_\delta\dot z^\gamma\dot z^\delta 
\nonumber \\ && \qquad
-\int^{\tau_x}_{\mp\infty}d\tau'
v^{\alpha\beta}{}_{\alpha'\beta'}(z(\tau_x),z(\tau'))
\dot z^{\alpha'}(\tau')\dot z^{\beta'}(\tau')
\nonumber \\ && \qquad
+\sigma_{;\gamma}\int^{\tau_x}_{\mp\infty}d\tau'
v^{\alpha\beta}{}_{\alpha'\beta'}{}^{;\gamma}
(z(\tau_x),z(\tau'))\dot z^{\alpha'}(\tau')\dot z^{\beta'}(\tau') \Biggr)
+O(\epsilon^2, \tau_r^{-1}\epsilon). \label{eq:metper0}
\end{eqnarray}
In the above and in what follows, we omit the suffix $Ret/Adv$ to 
$\psi_{\mu\nu}$ for notational simplicity.
The detailed derivation of the above formula is provided in Appendix A.
Hereafter we also suppress the arguments $x$, $z(\tau_x)$ and $\tau_x$
unless there arises ambiguity.

The covariant derivative of the metric perturbation becomes 
also necessary in section 3. 
For this purpose, we calculate the covariant derivatives 
of the quantities appearing in Eq.~(\ref{eq:metper0}) 
such as $\tau_x$, $\epsilon(x)$ and so on. 
With the aid of the relations given in Appendix A, 
the derivative of $\tau_x$ can be calculated  
by taking the derivative of the both sides of 
the identity (\ref{eq:foli}) as 
\begin{eqnarray}
\tau_{x;\mu}&=&
\kappa^{-2}\bar g_{\mu\alpha}\left(-\dot z^\alpha
+{1\over 6}R^\alpha{}_{\beta\gamma\delta}
\sigma^{;\beta}\dot z^\gamma\sigma^{;\delta}
\right) +O(\epsilon^3).
\label{diftaux}
\end{eqnarray}
One must be careful that 
$[f(x,z(\tau_x))]_{;\mu}\not=[f(x,z(\tau))_{;\mu}]_{\tau=\tau_x}$
since $\tau_x$ is $x$ dependent through Eq.~(\ref{eq:foli}).
By using Eq.~(\ref{diftaux}) and 
the relations given in Appendix A, we obtain
\begin{eqnarray}
\epsilon_{;\mu}(x)&=&
-{1\over \epsilon}\bar g_{\mu\alpha}\sigma^{;\alpha},
\\
\left[ \sigma^{;\alpha}(z(\tau_x),x) \right]_{;\mu}
&=&\bar g_{\mu\beta}
\Biggl(-g^{\alpha\beta}-\kappa^{-2}\dot z^\alpha\dot z^\beta 
\nonumber \\ && \qquad 
+{1\over 6}\biggl(g^{\alpha\gamma}g^{\beta\epsilon}
-2g^{\alpha\gamma}\dot z^\beta\dot z^\epsilon 
+\dot z^\alpha g^{\beta\gamma}\dot z^\epsilon \biggr)
R_{\gamma\delta\epsilon\zeta}\sigma^{;\delta}\sigma^{;\zeta} \Biggr)
+O(\epsilon^2),
\\
\left[ \bar g^{\mu\alpha}(x,z(\tau_x)) \right]_{;\nu}&=&
{1\over 2}\bar g^\mu{}_\beta\bar g_{\nu\gamma}
\left(g^{\gamma\delta}-\dot z^\gamma\dot z^\delta\right)
R^{\alpha\beta}{}_{\delta\epsilon}\sigma^{;\epsilon}+O(\epsilon^2), 
\\
\kappa_{;\nu}(x)
&=&{1\over 2\kappa} \left[-\ddot \sigma(x,z(\tau_x)) \right]_{;\nu}
\nonumber \\
&=&{1\over 2\kappa}\bar g_{\nu\alpha}
\left(\ddot z^\alpha +\dot z^\alpha\sigma_{;\beta}\stackrel{...}{z}{}^\beta
+{2\over 3} R^\alpha{}_{\beta\gamma\delta}\dot z^\beta\sigma^{;\gamma}
\dot z^\delta\right)
+O(\epsilon^3).
\end{eqnarray}
Then using these results and Eq.~(\ref{eq:ulim}), 
after some computations, we obtain the 1st and 2nd derivatives of 
the metric perturbation $\psi_{\mu\nu}(x)$. 
\begin{eqnarray}
\psi_{\mu\nu;\xi}(x)
&=&2Gm\bar g_{\mu\alpha}\bar g_{\nu\beta}\bar g_{\xi\gamma} 
\nonumber \\ && \quad 
\Biggl({2\over\epsilon^3}\kappa^{-1}\dot z^\alpha\dot z^\beta\sigma^{;\gamma}
-{1\over\epsilon}\dot z^\alpha\dot z^\beta\ddot z^\gamma 
-{4\over\epsilon}\dot z^{(\alpha}\ddot z^{\beta)}\dot z^\gamma 
\pm2\dot z^{(\alpha} 
R^{\beta)}{}_\delta{}^\gamma{}_\epsilon\dot z^{\delta}\dot z^\epsilon 
\mp2R^\alpha{}_{\delta}{}^\beta{}_\epsilon \dot z^\gamma\dot z^{\delta}
\dot z^\epsilon 
\nonumber \\ && \qquad 
-{2\over\epsilon}\dot z^{(\alpha} 
R^{\beta)}{}_\delta{}^\gamma{}_\epsilon\dot z^{\delta}\sigma^{;\epsilon}
+{2\over\epsilon}R^\alpha{}_\delta{}^\beta{}_\epsilon 
\sigma^{;\gamma}\dot z^{\delta}\dot z^\epsilon 
-{2\over\epsilon}\dot z^{(\alpha} R^{\beta)}{}_{\delta\epsilon\zeta}
\dot z^\gamma\dot z^{\delta}\sigma^{;\epsilon}\dot z^\zeta
-{2\over 3\epsilon}\dot z^\alpha \dot z^\beta 
R^\gamma{}_{\delta\epsilon\zeta}\dot z^{\delta}\sigma^{;\epsilon}\dot z^\zeta
\nonumber \\ && \qquad 
\mp\int^{\tau_x}_{\mp\infty}d\tau'
v^{\alpha\beta}{}_{\alpha'\beta'}{}^{;\gamma}(z(\tau_x),z(\tau'))
\dot z^{\alpha'}(\tau')\dot z^{\beta'}(\tau') 
\Biggr)
+O(\epsilon^1, \tau_r^{-1}),
\label{eq:metper1}
\\
\psi_{\mu\nu;\xi\rho}(x)
&=&2Gm\bar g_{\mu\alpha}\bar g_{\nu\beta}\bar g_{\xi\gamma}\bar g_{\rho\delta}
\nonumber \\ && \quad 
\Biggl(-{2\over\epsilon^3}\kappa^{-1}\dot z^\alpha\dot z^\beta
\left(g^{\gamma\delta}+\kappa^{-2}\dot z^\gamma\dot z^\delta 
-{3\over \epsilon^2}\sigma^{;\gamma}\sigma^{;\delta}\right)
-{2\over\epsilon^3}
\dot z^\alpha\dot z^\beta\ddot z^{(\gamma}\sigma^{;\delta)} 
-{8\over\epsilon^3}
\dot z^{(\alpha}\ddot z^{\beta)}\dot z^{(\gamma}\sigma^{;\delta)} 
\nonumber \\ && \qquad 
+{2\over\epsilon}\dot z^{(\alpha}R^{\beta)}
{}_\epsilon{}^{\gamma\delta}\dot z^\epsilon 
+{2\over3\epsilon}\dot z^\alpha\dot z^\beta 
R^\gamma{}_\epsilon{}^\delta{}_\zeta
\dot z^\epsilon \dot z^\zeta
+{4\over\epsilon}\dot z^{(\alpha}R^{\beta)}{}_\epsilon{}^{(\gamma}{}_\zeta
\dot z^{\delta)}\dot z^\epsilon\dot z^\zeta
\nonumber \\ && \qquad 
-{2\over\epsilon}R^\alpha{}_\epsilon{}^\beta{}_\zeta
\left(g^{\gamma\delta}+\dot z^\gamma\dot z^\delta 
-{1\over\epsilon^2}\sigma^{;\gamma}\sigma^{;\delta}\right)
\dot z^\epsilon\dot z^\zeta
-{4\over\epsilon^3}\dot z^{(\alpha}R^{\beta)}{}_\epsilon{}^{(\gamma}{}_\zeta
\sigma^{;\delta)}\dot z^\epsilon\sigma^{;\zeta}
\nonumber \\ && \qquad 
-{2\over 3\epsilon^3}\dot z^\alpha\dot z^\beta 
R^\gamma{}_\epsilon{}^\delta{}_\zeta
\sigma^{;\epsilon}\sigma^{;\zeta}
-{4\over\epsilon^3}\dot z^{(\alpha}R^{\beta)}{}_{\epsilon\zeta\eta}
\dot z^{(\gamma}\sigma^{;\delta)} \dot z^\epsilon\sigma^{;\zeta}\dot z^\eta
-{4\over3\epsilon^3}\dot z^\alpha\dot z^\beta\sigma^{;(\gamma}
R^{\delta)}{}_{\epsilon\zeta\eta}\dot z^\epsilon\sigma^{;\zeta}\dot z^\eta
\nonumber \\ && \qquad 
+{2\over 3\epsilon^3}\dot z^\alpha\dot z^\beta\dot z^{(\gamma}
R^{\delta)}{}_{\epsilon\zeta\eta}\sigma^{;\epsilon}
\dot z^\zeta\sigma^{;\eta} \Biggr)
+O\left(\epsilon^0, {\tau_r^{-1}\over\epsilon}\right).
\label{eq:metper2}
\end{eqnarray}
We note that among the terms on the right hand side of the
above expressions, the terms involving the Riemann tensor
will not contribute to the equation of motion, as will be shown 
later.

Using the above expressions, one can directly check that 
$\psi^{\mu\nu;\xi}{}_\xi+2 R^\mu{}_\xi{}^\nu{}_\rho\psi^{\xi\rho}=0$ 
is satisfied for $\epsilon \not= 0$ to the order in which 
we are concerned. 
To the contrary, the harmonic gauge condition is not automatically 
satisfied. 
We find 
\begin{equation} 
\psi^{\mu\nu}{}_{;\nu}={8Gm\over\epsilon} \bar g^{\mu}{}_{\alpha} 
\ddot z^{\alpha}.
\end{equation}
Generally speaking, thus obtained metric perturbation 
satisfies the harmonic gauge condition only when 
the source energy-momentum tensor satisfies 
the divergence free condition,  
which, in the present point-like particle case, 
leads to the conclusion that the source trajectory 
must be a geodesic. 
This point is totally different from the electromagnetic counterpart, 
in which the Lorentz gauge condition is related to the electric 
charge conservation. 
However this does not mean the break down of our formalism. 
Since $\ddot z$ is assumed to be a higher order quantity, 
the missing of the harmonic condition 
is responsible for the neglection of the higher 
order perturbations.

\section{DeWitt and Brehme's Approach}
In this section, we develop our discussion in an analogous 
way to the 
electromagnetic counterpart given by DeWitt and Brehme. 
First we derive the `conserved' rank two symmetric tensor. 
Integrating its divergence over the interior of the world 
tube surrounding the particle, we derive the equation of 
motion including the effect of the gravitational radiation 
reaction.

\subsection{Conservation law} 

In the formalism developed by DeWitt and Brehme, 
crucially important was 
the conserved energy-momentum tensor, which 
consists of the matter and field contributions. 
In the case of gravity, the matter energy-momentum tensor 
is divergence free by itself in the sense of the covariant 
derivative with respect to the total metric, $\tilde g_{\mu\nu}(x)$. 
Thus the situation looks different. 
However, if we choose the background spacetime $g_{\mu\nu}(x)$, 
we can construct a quantity analogous to the 
conserved energy-momentum tensor in the electromagnetic case. 

We divide the metric $\tilde g_{\mu\nu}(x)$ 
into the background and the deviation from it, 
$\delta g_{\mu\nu}(x)$, as 
\begin{eqnarray}
\tilde g_{\mu\nu}(x)=g_{\mu\nu}(x)+\delta g_{\mu\nu}(x). 
\label{eq:full}
\end{eqnarray}
The background metric is assumed to be 
a solution of the vacuum Einstein equations. 
We write the Einstein equations with matter source, 
whose degrees of freedom are represented by ${\bf\phi}$ symbolically, 
as 
\begin{equation}
G^{\mu\nu}[{\bf g}+\delta{\bf g}]=
8\pi G T^{\mu\nu}[{\bf g}+\delta{\bf g},{\bf\phi}].
\end{equation}
We expand $G^{\mu\nu}[{\bf g}+\delta{\bf g}]$ and
$T^{\mu\nu}[{\bf g}+\delta{\bf g},{\bf\phi}]$ with respect 
to $\delta g_{\mu\nu}$ as 
\begin{eqnarray}
&&G^{\mu\nu}[{\bf g}+\delta{\bf g}]=
G^{(0)\mu\nu}+G^{(1)\mu\nu}[\delta{\bf g}]+G^{(2+)\mu\nu}[\delta{\bf g}],
\nonumber \\\\ 
&&T^{\mu\nu}[{\bf g}+\delta{\bf g},{\bf\phi}]=
T^{(0)\mu\nu}[{\bf\phi}]+T^{(1+)\mu\nu}[\delta{\bf g},{\bf\phi}], 
\end{eqnarray}
where the superscript $(n)$ represents the terms of the $n$th order 
in the metric perturbation $\delta{\bf g}$ and 
the superscript $(n+)$ 
represents the $n$th or higher order terms.
Then the Einstein equations are rewritten as
\begin{equation}
{\cal T}^{\mu\nu}[\delta{\bf g},{\bf\phi}]:=
T^{(0)\mu\nu}[{\bf\phi}]+T^{(1+)\mu\nu}[\delta{\bf g},{\bf\phi}]
 -{1\over 8\pi G}G^{(2+)\mu\nu}[\delta{\bf g}]
={1\over 8\pi G}G^{(1)\mu\nu}[\delta{\bf g}]. 
\label{eq:ct} 
\end{equation} 
{}From the $O(\delta {\bf g})$ terms of the contracted Bianchi 
identities, we find $G^{(1)\mu\nu}{}_{;\nu}[\delta{\bf g}]=0$ 
when the background is a solution of the vacuum
Einstein equations. Here we note again that the semicolon means the 
covariant derivative with respect to the background metric. 
Thus we obtain the covariant conservation law, 
\begin{eqnarray}
{\cal T}^{\mu\nu}{}_{;\nu}[\delta{\bf g},{\bf\phi}]=0,
\end{eqnarray}
which is what we needed.

One can see that 
if the mass of the particle is small enough and so 
the metric perturbation $\delta g_{\mu\nu}(x)$ is negligibly small, 
the conserved tensor ${\cal T}^{\mu\nu}[\delta{\bf g},{\bf\phi}]$ reduces to 
$T^{(0)\mu\nu}[{\bf\phi}]$ which is independent of $\delta{\bf g}$, 
which implies $T^{(0)\mu\nu}{}_{;\nu}[{\bf\phi}]=0$. 
Thus, in the lowest order of the metric perturbation, 
one observes that the point particle 
moves along the background geodesic. 

We shall now specify our consideration to the case in 
which $\delta g_{\mu\nu}(x)$ is so small that  
we can replace $\delta g_{\mu\nu}$ by 
the linear perturbation $h_{\mu\nu}$ induced by 
$T^{(0)\mu\nu}[{\bf\phi}]$. 
Then the conserved tensor ${\cal T}^{\mu\nu}$ can be approximated as 
\begin{eqnarray}
{\cal T}^{\mu\nu}[\delta{\bf g},{\bf\phi}]\approx{\cal T}^{\mu\nu}[{\bf h},\phi]
=T^{(0)\mu\nu}[\phi]+T^{(1)\mu\nu}[{\bf h},{\bf\phi}]
-{1\over 8\pi G}G^{(2)\mu\nu}[{\bf h}].
\end{eqnarray}
The 1st and 2nd terms on the right hand side of the equation
vanish outside the matter distribution, 
while the 3rd term bilinear in ${\bf h}$ does not 
vanish anywhere. 
Therefore it may be interpreted as the gravitational 
contribution to the energy-momentum tensor. 
The 3rd term is expressed in terms of
the trace-reversed metric perturbation $\psi_{\mu\nu}$ as
\begin{eqnarray}
G^{(2)\mu\nu}[{\bf h}]&=&{1\over2}\psi\, G^{(1)\mu\nu}[{\bf h}] 
\nonumber \\ && 
-{1\over 2}\Biggl\{{1\over 2}
\biggl(\psi^{\mu\xi; \rho}+\psi^{\mu\rho;\xi}-\psi^{\xi\rho;\mu}\biggr)
\biggl(\psi^\nu{}_{\xi; \rho}+\psi^\nu{}_{\rho;\xi}
-\psi_{\xi\rho}{}^{;\nu}\biggr)
\nonumber \\ && \qquad
+\psi^{\mu\xi}{}_{;\rho\xi}\psi^{\nu\rho}
+\psi^{\nu\xi}{}_{;\rho\xi}\psi^{\mu\rho}
-\psi^{\mu\nu}{}_{;\rho\xi}\psi^{\rho\xi}-{1\over 4}\psi^{;\mu}\psi^{;\nu}
\nonumber \\ && \qquad
+{1\over 4}g^{\mu\nu}\biggl(-\psi^{\xi\rho;\sigma}\psi_{\xi\rho;\sigma}
+2\psi^{\xi\rho;\sigma}\psi_{\rho\sigma;\xi}+{1\over 2}\psi^{;\xi}\psi_{;\xi}
\biggr)\Biggr\}.
\label{eq:2ndein}
\end{eqnarray}
The derivation of the above formula 
is given in the Appendix B. 

\subsection{World Tube Around The Particle}
Following DB, we introduce a further parameterization which
distinguishes the points on the same 3-surface parameterized by $\tau$.
It is defined by an implicit relation between $x$ and the 4 parameters,
 i.e., $\tau_x$, $\epsilon$, and $\Omega^i$ $(i=1,2,3)$, 
\begin{eqnarray}
\sigma_{;\alpha}(x,z(\tau_x))=-\epsilon n_{\alpha i}(\tau_x)\Omega^i, 
\end{eqnarray}
where $\displaystyle \sum_{i=1}^3\Omega^i\Omega^i=1$. 
Here $n_{\alpha i}(\tau_x)$ is a set of orthonormal basis 
on the 3-hypersurface of $\tau_x$ at $z(\tau_x)$. 
It is defined by
\begin{eqnarray}
n_{\alpha i}(\tau)\dot z^\alpha(\tau)&=&0, \\
g^{\alpha\beta}(z(\tau))n_{\alpha i}(\tau)n_{\beta j}(\tau)&=&\delta_{ij}, 
\end{eqnarray}
at some $\tau_x$ and Fermi-Walker transported along the trajectory. 
In the following, 
we also use $\Omega^\alpha:=n^\alpha{}_i\Omega^i$. 
The world tube of the particle is defined in the same manner as in DB.
We consider the 4-volume of the interior of the world tube 
between two 3-surfaces of $\tau_{1}$ and $\tau_{2}$,  
\begin{eqnarray}
V_{tube}=\left\{x=x(\tau,\epsilon,\Omega^i);\ 
\tau_{1}\leq\tau\leq\tau_{2},\ 
0\leq\epsilon\leq\epsilon_{tube}\right\}.
\end{eqnarray}
We define the surfaces of this volume, 
\begin{eqnarray}
\Sigma_{tube}&=&\left\{x=x(\tau,\epsilon_{tube},\Omega^i);\ 
\tau_{1}\leq\tau\leq\tau_{2}\right\}, 
\\
\Sigma_{cap}(\tau)&=&\left\{x=x(\tau,\epsilon,\Omega^i);\ 
0\leq\epsilon\leq\epsilon_{tube}\right\}. 
\end{eqnarray}
The volume measure is given by $dV=\sqrt{-g(x)}d^4x$.
The integral measures on $\Sigma_{tube}$ and $\Sigma_{cap}(\tau)$ 
have been already derived in DB,
\begin{eqnarray}
d\Sigma_\mu(x)|_{\Sigma_{tube}}&=&\left.
\left( \epsilon^2 \kappa^2(x)
\bar g_{\mu\alpha}(x,z(\tau))\Omega^\alpha 
+O(\epsilon^5) \right)\right\vert_{\Sigma_{tube}}
d^2\Omega ~d\tau \,,
\\ 
d\Sigma_\mu(x)|_{\Sigma_{cap}(\tau)}&=&\left.
{1\over \Delta(x,z(\tau))}\epsilon^2
\sigma_{;\mu\alpha}(x,z(\tau))\dot z^\alpha(\tau)\right\vert_{\Sigma_{cap}(\tau)} 
~d\epsilon ~d^2\Omega\, ,
\end{eqnarray}
(See Eqs.~(4.35) and (4.45) of DB with $R^{\alpha\beta}(z)=0$. 
See also Eq.~(1.51) of DB. Note that our measure is 
with respect to the vector, 
while DB defines it with respect to the vector density.). 

Now integrating the null quantity ${\cal T}^{\mu\nu}{}_{;\nu}(x)$ 
over $V_{tube}$ and using the integration by part, we obtain  
\begin{eqnarray}
0&=&\int_{V_{tube}}dV~ 
\bar g^{\bar\alpha}{}_\beta (z(\bar\tau),z(\tau_x)) 
\bar g^\beta{}_\mu(z(\tau_x),x) 
{\cal T}^{\mu\nu}{}_{;\nu}(x)
\nonumber \\ 
&=&\int_{\Sigma_{tube}+\Sigma_{cap}(\tau_2)-\Sigma_{cap}(\tau_1)}
d\Sigma_\nu(x) \bar g^{\bar\alpha}{}_\beta (z(\bar\tau),z(\tau_x)) 
\bar g^\beta{}_\mu(z(\tau_x),x) 
{\cal T}^{\mu\nu}(x)
\nonumber \\ && \qquad \qquad 
-\int_{V_{tube}}dV
\left[\bar g^{\bar\alpha}{}_\beta (z(\bar\tau),z(\tau_x)) 
\bar g^\beta{}_\mu(z(\tau_x),x) 
\right]_{;\nu}{\cal T}^{\mu\nu}(x),
\label{eq:integ}
\end{eqnarray}
where $\bar\tau=(\tau_{1}+\tau_{2})/2$.
The reason why $\bar g^{\bar\alpha}{}_\beta (z(\bar\tau),z(\tau_x)) 
\,\bar g^\beta{}_\mu(z(\tau_x),x) $ is multiplied 
is to make the integrand to be a vector at $z(\bar\tau)$ and to behave 
as a scalar with respect to $x$, 
so that the integration is done in the covariant manner.

\subsection{Gravitational Radiation Damping}
To evaluate the second and third lines of Eq.~(\ref{eq:integ}), 
we have to put an assumption on the matter configuration. 
We take the tube radius $\epsilon_{tube}$ so that 
the tube is in the external zone. 
Therefore, the metric perturbation on the tube surface $\Sigma_{tube}$ 
can be approximated by that induced by a point-like particle given in 
Eq.~(\ref{eq:metper}). 
Noting that no contribution of 
matter to ${\cal T}^{\mu\nu}[{\bf h},\phi]$ exists 
there, ${\cal T}^{\mu\nu}[{\bf h},\phi]$ can be computed as
\begin{eqnarray}
{\cal T}^{\mu\nu}[{\bf h},\phi]
&=&{Gm^2\over 4\pi}\bar g^\mu{}_\alpha (x,z(\tau_x)) 
\bar g^\nu{}_\beta (x,z(\tau_x))
\nonumber \\ && 
\Biggl\{{1\over\epsilon^4}
\biggl(-4\dot z^\alpha\dot z^\beta
+\Omega^\alpha\Omega^\beta-{1\over 2}g^{\alpha\beta}\biggr)
+{1\over\epsilon^3}
\biggl(-4\dot z^\alpha\dot z^\beta\ddot z^\gamma
\Omega_\gamma-7\ddot z^{(\alpha}\Omega^{\beta)}
+{7\over 2}g^{\alpha\beta}\ddot z^\gamma\Omega_\gamma\biggr)
\nonumber \\ && \quad 
-{1\over\epsilon^2}\biggl(
8\dot z^{(\alpha}V^{\beta)}{}_{\gamma\delta}\dot z^{(\gamma}\Omega^{\delta)}
-4V_{\gamma\delta}{}^{(\alpha}\dot z^{\beta)}\dot z^\gamma\Omega^\delta 
+2g^{\alpha\beta}V_{\gamma\delta\epsilon}
\dot z^\gamma\Omega^\delta\dot z^\epsilon 
-4\Omega^{(\alpha}V^{\beta)}{}_{\gamma\delta}\dot z^\gamma\dot z^\delta 
\nonumber \\ && \qquad \qquad 
+2V_{\gamma\delta}{}^{(\alpha}\Omega^{\beta)}\dot z^\gamma\dot z^\delta 
-g^{\alpha\beta}V_{\gamma\delta\epsilon}\dot z^\gamma
\dot z^\delta\Omega^\epsilon 
+\Omega^{(\alpha}V^{\beta)}
-{1\over 2}g^{\alpha\beta}V_\gamma\Omega^\gamma\biggr)\Biggr\}
\nonumber \\ && \qquad \qquad \quad 
+{\cal T}^{\mu\nu}_R +O\left(
\stackrel{...}{z},{\tau_r^{-1}\over \epsilon^2},
{1\over\epsilon}\right), 
\label{eq:consten}
\end{eqnarray}
where $z,\dot z$, $\ddot z$ are evaluated at $\tau=\tau_x$,
 and we have defined
\begin{eqnarray} 
V_{\mu\nu\xi}(x)&:=&\mp\int^{\tau_x}_{\mp\infty}d\tau' 
v_{\mu\nu\alpha'\beta';\xi}(x,z(\tau'))
\dot z^{\alpha'}(\tau')\dot z^{\beta'}(\tau'), 
\\ 
V_{\mu}(x)&:=&\mp\int^{\tau_x}_{\mp\infty}d\tau' 
v^{\nu}{}_{\nu\alpha'\beta';\mu}(x,z(\tau'))
\dot z^{\alpha'}(\tau')\dot z^{\beta'}(\tau'). 
\end{eqnarray} 
Here we have denoted the terms which 
contain the Riemann tensor by ${\cal T}^{\mu\nu}_R$, 
which is at most of $O(1/\epsilon^2)$. 
It is not necessary to write them down explicitly because 
they do not affect the 
equation of motion, as will be shown later. 

Noting that the terms containing odd number of $\Omega$'s 
in the integrand vanish when $d^2\Omega$-integration is done, 
and that $\int d\Omega^2 \Omega^\alpha\Omega^{\beta}
={4\pi\over 3}\left(g^{\alpha\beta}+\dot z^\alpha\dot z^\beta\right)$, 
the integration over the surface of the tube $\Sigma_{tube}$ 
for small $\delta\tau:=\tau_{2}-\tau_{1}$ is evaluated as 
\begin{eqnarray}
&& \int_{\Sigma_{tube}}d\Sigma_\nu(x)
\bar g^{\bar\alpha}{}_\beta(z(\bar\tau),z(\tau_x)) 
\bar g^\beta{}_\mu(z(\tau_x),x) {\cal T}^{\mu\nu}[{\bf h}](x)
\nonumber \\ &&  
=Gm^2\Biggl\{\Biggl(
-{7\over 2\epsilon_{tube}}\ddot z^{\bar\alpha}
-{2\over 3}\dot z^{\bar\alpha} 
  V_{\bar\beta \bar\gamma \bar\delta}
\dot z^{\bar\beta}\dot z^{\bar\gamma}\dot z^{\bar\delta}
-{2\over 3}\dot z^{\bar\alpha} V_{\bar\beta}\dot z^{\bar\beta}
\nonumber \\ && \qquad \qquad \qquad 
+\left(V^{\bar\alpha}{}_{\bar\beta \bar\gamma}
+V^{\bar\alpha}{}_{\bar\gamma \bar\beta}
-V_{\bar\beta \bar\gamma}{}^{\bar\alpha}\right)
\dot z^{\bar\beta}\dot z^{\bar\gamma}
-{1\over 2}V^{\bar\alpha}\Biggr)(\bar\tau)
+O\left(\tau_r^{-1},\epsilon_{tube}\right)\Biggr\}\delta\tau
+O(\delta\tau^2).
\end{eqnarray}

The surface integration over the cap $\Sigma_{cap}$ requires 
the knowledge of the metric in the internal zone. 
We assume that the matter configuration and the metric is
not perturbed much and is kept spherically symmetric  
inside the radius $\epsilon_0(\gg Gm)$ to sufficient accuracy. 
In other words, locally the particle behaves as if it were an
isolated object.
Mathematically, we assume that 
the cap integration for $\epsilon\leq\epsilon_0$, 
which in an unrigorous manner defines
the particle momentum $p^\alpha_{\epsilon_0}(\tau)$, 
is proportional to 
$\dot z^\alpha(\tau)$, 
\begin{eqnarray}
\int_{\Sigma_{cap}(\tau),\epsilon<\epsilon_0}
&&d\Sigma_\nu(x)
\bar g^{\bar\alpha}{}_\beta(z(\bar\tau),z(\tau_{x})) 
\bar g^\beta{}_\mu(z(\tau_{x}),x)
{\cal T}^{\mu\nu}[{\bf h}](x)
\cr &&
= m(\epsilon_0,\tau) 
\bar g^{\bar\alpha}{}_\beta(z(\bar\tau),z(\tau)) \dot z^\beta(\tau)
+{G m^2} O\left(m,\tau_r^{-1},\epsilon_{tube}\right). 
\label{capin}
\end{eqnarray}
We call the matter which satisfies this assumption 
as an ideal point-like particle.

Outside the radius $\epsilon_0$, the metric is 
approximated by that induced by a point-like particle, (\ref{eq:metper}). 
Then the cap integration for $\epsilon_0\leq\epsilon\leq\epsilon_{tube}$ 
is evaluated by using Eq.~(\ref{eq:consten}) to give 
\begin{eqnarray}
\int_{\Sigma_{cap}(\tau),\epsilon>\epsilon_0}
&&d\Sigma_\nu(x)
\bar g^{\bar\alpha}{}_\beta(z(\bar\tau),z(\tau_{x}))
\bar g^\beta{}_\mu(z(\tau_x),x) {\cal T}^{\mu\nu}[{\bf h}](x)
\cr
&&=-{7\over 2}{Gm^2}\left[
\left({1\over \epsilon_0}-{1\over \epsilon_{tube}}\right)
\bar g^{\bar\alpha}{}_\beta(z(\bar\tau),z(\tau)) \dot z^{\beta}(\tau)
+O(\tau_r^{-1},\epsilon_{tube})\right] .
\label{capout}
\end{eqnarray}
Thus setting
\begin{equation}
 m(\epsilon_0,\tau)=m(\tau)+{7\over 2}{Gm^2\over\epsilon_0}, 
\end{equation}
we obtain
\begin{eqnarray}
\int_{\Sigma_{cap}(\tau_2)-\Sigma_{cap}(\tau_1)}
&&d\Sigma_\nu(x)
\bar g^{\bar\alpha}{}_\beta(z(\bar\tau),z(\tau_{x}))
\bar g^\beta{}_\mu(z(\tau_x),x) {\cal T}^{\mu\nu}[{\bf h}](x)
\cr
&&=\left[\left\{m(\bar\tau)+{7\over 2}{Gm^2\over\epsilon_{tube}}\right\}
\ddot z^{\bar\alpha}(\bar\tau) 
+\dot m(\bar\tau) \dot z^{\bar\alpha}(\bar\tau)\right]\delta\tau
+O(\delta\tau^2). 
\end{eqnarray} 

Finally, we consider the remaining volume integral
$\int_{V_{tube}}dV
[\bar g^{\bar\alpha}{}_{\beta} \bar g^\beta{}_{\mu}]_{;\nu} {\cal T}^{\mu\nu}$.
Since $[\bar g^{\bar\alpha}{}_{\beta}\bar g^\alpha{}_{\mu}]_{;\nu}
\approx \epsilon\Omega$, again we expect that 
the integration for $\epsilon<\epsilon_0$ 
vanishes for the ideal point-like particle 
to sufficient accuracy. 
The integration for $\epsilon>\epsilon_0$ is at most of
$O\left(Gm^2\epsilon_0 \delta\tau\right)$. 
Thus we may neglect the contribution from the volume integral.  

Putting all results together, 
we obtain the equation of motion,  
\begin{eqnarray}
 m(\tau) \ddot z^\alpha+\dot m(\tau) \dot z^{\alpha}
&= &Gm^2\Biggl\{
{2\over 3}\dot z^\alpha V_{\beta\gamma\delta}
\dot z^\beta\dot z^\gamma\dot z^\delta
+{2\over 3}\dot z^\alpha V_\beta\dot z^\beta
\nonumber \\ && \qquad  
-\left(V^{\alpha}{}_{\beta\gamma}+V^{\alpha}{}_{\gamma\beta}
-V_{\beta\gamma}{}^\alpha\right)\dot z^\beta\dot z^\gamma
+{1\over 2}V^{\alpha}+O(\epsilon_{tube},\tau_r^{-1})\Biggr\}. 
\label{eq:unnorm}
\end{eqnarray}
The leading contributions from the 
terms proportional to the Riemann tensor which we have neglected 
are at most of the zeroth order in $\epsilon_{tube}$ (or $\epsilon_{0}$). 
As they should be 
vectors, they must take the form 
$Gm^2 R^{\mu}{}_{\nu\rho\sigma}\dot z^{\nu}\dot z^{\rho}\dot z^{\sigma}$ 
or $Gm^2 R^{\mu}{}_{\nu\rho\sigma} g^{\nu\rho}\dot z^{\sigma}$
by means of the dimensional argument. 
Thus all the leading terms disappear and the remaining terms become 
higher order in $\epsilon_{tube}$ (or $\epsilon_{0}$). 

Now we consider the normalization condition 
$g_{\alpha\beta}(z(\tau))\dot z^\alpha(\tau)\ddot z^\beta(\tau)=0$. 
This gives 
\begin{equation}
 \dot m(\tau)=Gm^2\Biggl\{
{5\over 3} V_{\beta\gamma\delta}\dot z^\beta\dot z^\gamma\dot z^\delta
+{1\over 6} V_\beta\dot z^\beta\Biggr\},
\label{mtaudot}
\end{equation}
and it can be integrated as
\begin{equation}
   m(\tau)=m\Biggl\{1+
\left({5\over 6} \dot z^\beta\dot z^\gamma
+{1\over 12} g^{\beta\gamma}\right) \psi_{(v)}{}_{\beta\gamma}
\left(z(\tau)\right)\Biggr\} , 
\label{mtau}
\end{equation}
where we have defined 
\begin{equation}
\psi_{(v)}{}_{\mu\nu}(x):=\mp2Gm\int^{\tau_x}_{\mp\infty}d\tau' 
v_{\mu\nu\alpha'\beta'}(x,z(\tau'))
\dot z^{\alpha'}(\tau')\dot z^{\beta'}(\tau'), 
\label{eq:psiv}
\end{equation}
which is the part of the trace-reversed metric perturbation due
 to the $v_{\mu\nu\alpha\beta}(x,z)$ term (i.e, the so-called tail term)
in the Green function (see Eq.~(\ref{eq:metper})). 
Since $\psi_{(v)\beta\gamma}$ is expected to be of $O(Gm)$, 
Eq.~(\ref{mtau}) tells us that we can consistently replace 
$m(\tau)$ by $m$ in Eq.~(\ref{eq:unnorm}) except for the term, 
$\dot m(\tau) \dot z^{\alpha}$. 

Substituting Eq.~(\ref{mtaudot}) into Eq.~(\ref{eq:unnorm}), 
we finally obtain 
\begin{eqnarray}
 m\,\ddot z^\alpha(\tau)
 = -m\Biggr(&&
{1\over 2}\dot z^\alpha\dot z^\beta\dot z^\gamma\dot z^\delta 
+g^{\alpha\beta}(z) \dot z^\gamma\dot z^\delta 
-{1\over 2}g^{\alpha\delta}(z)\dot z^\beta\dot z^\gamma 
\cr &&
-{1\over 4}\dot z^\alpha g^{\beta\gamma}(z)\dot z^\delta 
-{1\over 4}g^{\alpha\delta}(z)g^{\beta\gamma}(z)\Biggl)(\tau)
\,\psi_{(v)}{}_{\beta\gamma;\delta}(z(\tau)), 
\label{eq:damp}
\end{eqnarray}
where we have also used the relation 
$V_{\alpha\beta\gamma}(z(\tau))=\psi_{(v)}{}_{\alpha\beta;\gamma}(z(\tau))$ 
which follows from Eq.~(\ref{eq:ulim}). 
If we impose the physical boundary condition 
with no-incoming waves from the past null infinity, 
we should take the upper sign (i.e.,
the retarded boundary condition) of
Eq.~(\ref{eq:psiv}) for $\psi_{(v)\beta\gamma}$.
The meaning of this equation is discussed in section 5. 

\section{Matched Asymptotic Expansion}
In this section we give an alternative derivation of the 
equation of motion obtained in the previous section 
in a more rigorous way by using the 
matched asymptotic expansion technique. 

\subsection{Matching Scheme}
To begin with, we state the general concept of 
the asymptotic matching. 
We first prepare the metrics in both internal and external zones 
by using different schemes. 
In the internal zone, we expect that the metric can be described 
by using the black hole perturbation. 
Namely, we assume that the particle is represented by 
a Schwarzschild black hole in the lowest order of 
approximation.
In the present case, 
the perturbation is caused by the tidal effects of the curvature of the 
spacetime in which the particle travels. 
We call this construction of the metric the internal scheme. 
In order to make this scheme valid, the linear extension of
the internal zone around the particle 
must be much smaller than the background curvature scale. 
We use the coordinate 
$\{ X^a \}=\{ T, X^i \}\quad (a=0,1,2,3;~i=1,2,3)$ 
for the internal scheme and $|X|(:=\sqrt{X^iX^i})$ is assumed to
represent the physical distance scale
(in this section, we adopt
the Minkowskian summation rule on $a,b,\cdots$, 
and the Kronecker summation rule on $i,j,\cdots$ 
over the repeated indices).
In the external zone, as discussed in the previous sections, 
we expect that the metric is represented by the perturbations 
induced by a particle on a given background spacetime. 
We call this construction of the metric the external scheme. 
We require that the metrics obtained in both schemes be matched 
in the matching region of both zones, 
by considering the coordinate transformation 
between the internal and external zones. 
Safely, we may assume the existence of the matching region 
as long as $Gm\ll L\sim1$. 
we set the matching radius at $|X| \sim(GmL)^{1/2}$ by 
using the spatial coordinates of the internal scheme, {$X^i$}. 
Then, writing down the metric in the internal scheme, 
we have two independent small parameters $|X|/L$ and $Gm/|X|$ 
in the matching region. 
The power expansion with respect to these two small parameters 
allows us to consider the matching order by order. 

First we consider the expansion of the internal scheme. 
Recalling that the perturbation in the internal zone is 
induced by the external curvature 
which is characterized by the length $L$, 
the metric can be expanded in powers of $|X|/L$ as 
\begin{equation}
\tilde g_{ab}(X) = 
\mathop{H}\limits^{(0)}{}_{ab}(X)+{1\over L}\mathop{H}^{(1)}{}_{ab}(X)
+{1\over L^2}\mathop{H}^{(2)}{}_{ab}(X)+\cdots, 
\label{eq:bh0}
\end{equation}
where $\mathop{H}\limits^{(0)}{}_{ab}(X)$ represents the 
unperturbed black hole metric. 
We expect that 
$\mathop{H}\limits^{(1)}{}_{ab}(X)$ will be given by the 
standard linear perturbation 
of the black hole. 
Later, we find that the matching condition 
requires that $\mathop{H}\limits^{(1)}{}_{ab}(X)$ should vanish, 
which is consistent with the notion that 
the spacetime curvature is of $O(1/L^2)$. 
Thus the standard black hole perturbation theory applies 
up to $\mathop{H}\limits^{(2)}{}_{ab}(X)$. 
As is known well, the linear perturbations of the Schwarzschild 
black hole \cite{Zerilli} 
can be decomposed by using the tensor harmonics, 
which are classified by the total angular momentum, $J$. 
The monopole mode $(J=0)$ corresponds 
to the mass perturbation. 
Thus we may set this mode to vanish 
since it is natural to suppose that 
the change of mass due to the radiation reaction 
is small. 
The dipole modes $(J=1)$ are related to the 
translation and rotation. 
The translation modes are purely gauge and thus 
we set them to vanish to fix the center of the black hole. 
As we are considering a non-rotating black hole, 
we also set the rotational modes to vanish. 
In general, the higher modes contain
gauge degrees of freedom as well as the physical ones. 
However, for these higher modes, 
we do not give any principle to fix the gauge for the moment.
Further we expand the metric with respect to $Gm/|X|$ as
\begin{eqnarray}
\mathop{H}^{(0)}{}_{ab}(X)&=&\eta_{ab}+Gm\mathop{H}_{(1)}^{(0)}{}_{ab}(X)
+(Gm)^2 \mathop{H}_{(2)}^{(0)}{}_{ab}(X) 
+\cdots\,,
\nonumber \\ 
{1\over L}\mathop{H}^{(1)}{}_{ab}(X)&=&
{1\over L}\mathop{H}_{(0)}^{(1)}{}_{ab}(X)
+{Gm\over L}\mathop{H}_{(1)}^{(1)}{}_{ab}(X)
+{(Gm)^2\over L}\mathop{H}_{(2)}^{(1)}{}_{ab}(X)+\cdots\,,
\nonumber \\ 
{1\over L^2}\mathop{H}^{(2)}{}_{ab}(X)&=&
{1\over L^2}\mathop{H}_{(0)}^{(2)}{}_{ab}(X)
+{Gm\over L^2}\mathop{H}_{(1)}^{(2)}{}_{ab}(X)
+{(Gm)^2\over L^2}\mathop{H}_{(2)}^{(2)}{}_{ab}(X)+\cdots\,.
\label{eq:bh}
\end{eqnarray}
Note that, from the definitions of the expansion parameters,
the component of the metric behaves as 
\begin{eqnarray}
\mathop{H}_{(n)}^{(m)}{}_{ab} \sim |X|^{(m-n)} 
\label{eq:bhpower}. 
\end{eqnarray}

In order to write down the external metric in terms of the
internal coordinates, 
we consider the coordinate transformation from $x$ to $(T,X^i)$ given
in the form,
\begin{eqnarray}
\sigma_{;\alpha}(x,z(T))=-{\cal F}_\alpha(T,X) \label{eq:trans}. 
\end{eqnarray}
We assume $X^i=0$ corresponds to the center of the particle, 
$x^\alpha=z^\alpha(T)$, hence ${\cal F}_\alpha=0$ at $X^i=0$. 
We suppose that the right hand side of Eq.~(\ref{eq:trans}) can 
be expanded in positive powers of $X^i$ as 
\begin{eqnarray}
{\cal F}_\alpha(T,X)={f}_{\alpha i}(T)X^i+{1\over 2}{f}_{\alpha ij}(T)X^iX^j
+{1\over 3!}{f}_{\alpha ijk}(T)X^iX^jX^k+\cdots.
\end{eqnarray}
Although it is possible that 
more complicated terms such as $X^iX^j/|X|$ may appear,
we simply ignore these kinds of terms.
We shall find it is consistent 
within the order of the approximation to 
which we are going to develop our consideration below. 
Here ${f}_{\alpha i_1 \cdots i_n}(T)$ is totally 
symmetric for $i_1\cdots i_n$ 
and is at most of $O(L^{-(n-1)})$.
Using Eqs.~(\ref{128db}) and (\ref{173db}) in Appendix A, the
total derivative of Eq.~(\ref{eq:trans}) gives the important relation, 
\begin{eqnarray}
\bar g^\alpha{}_\mu(z(T),x)dx^\mu &=&
\Biggl( {dz^\alpha \over dT}(T) +{D {f}^{\alpha}{}_{i}\over dT}(T)X^i
+{1\over 2}{D {f}^{\alpha}{}_{ij}\over dT}(T)X^iX^j
\nonumber \\ && \qquad \quad
+{1\over 2}R^\alpha{}_{\beta\gamma\delta}(z(T))
{f}^{\beta}{}_{i}(T){d z^\gamma\over dT}(T)
{f}^{\delta}{}_{j}(T)X^iX^j+O(|X|^3)\Biggr) dT
\nonumber \\ &&
+\Biggl( {f}^{\alpha}{}_{i}+{f}^{\alpha}{}_{ij}(T)X^j
+{1\over 2}{f}^{\alpha}{}_{ijk}(T)X^jX^k
\nonumber \\ && \qquad \quad
+{1\over 6}R^\alpha{}_{\beta\gamma\delta}(z(T))
{f}^{\beta}{}_{j}(T){f}^{\gamma}{}_{i}(T)
{f}^{\delta}{}_{k}(T)X^jX^k+O(|X|^3)\Biggr) dX^i. 
\label{eq:dtrans}
\end{eqnarray}
Then, with the aid of Eqs.~(\ref{eq:trans}) and (\ref{eq:dtrans}), 
the external metric $\tilde g_{\mu\nu}(x)$ can be transformed into 
that written in terms of the internal coordinates by the relation, 
\begin{equation}
\tilde g_{ab}(X)dX^a dX^b=\tilde g_{\mu\nu}(x)dx^{\mu} dx^{\nu}.
\label{eq:465}
\end{equation}

Generally, as the external metric can be expanded by $Gm/|X|$, 
we write it as 
\begin{eqnarray}
 \tilde g_{ab}(X)= g_{ab}(X) +Gm \mathop{h}_{(1)}{}_{ab}(X)
+(Gm)^2 \mathop{h}_{(2)}{}_{ab}(X) +\cdots. 
\end{eqnarray}
Then $Gm \mathop{h}\limits_{(1)}{}_{ab}(X)$ can be recognized as 
the linear perturbation on the background $g_{ab}(X)$. 
Further we expand it with respect to $|X|/L$ as
\begin{eqnarray}
g_{ab}(X) &=& 
\mathop{h}^{(0)}_{(0)}{}_{ab}(X)+{1\over L}\mathop{h}^{(1)}_{(0)}{}_{ab}(X)
+{1\over L^2}\mathop{h}^{(2)}_{(0)}{}_{ab}(X)+\cdots\,,
\nonumber \\
Gm \mathop{h}_{(1)}{}_{ab}(X) &=& 
Gm\mathop{h}_{(1)}^{(0)}{}_{ab}(X)+{Gm\over L}\mathop{h}_{(1)}^{(1)}{}_{ab}(X)
+{Gm\over L^2}\mathop{h}_{(1)}^{(2)}{}_{ab}(X)+\cdots\,,
\nonumber \\ 
(Gm)^2 \mathop{h}_{(2)}{}_{ab}(X) &=& 
(Gm)^2\mathop{h}_{(2)}^{(0)}{}_{ab}(X)+{(Gm)^2\over L}
\mathop{h}_{(2)}^{(1)}{}_{ab}(X)
+{(Gm)^2\over L^2}\mathop{h}_{(2)}^{(2)}{}_{ab}(X) +\cdots\,. 
\label{eq:ext}
\end{eqnarray}
As before, 
\begin{eqnarray}
\mathop{h}_{(n)}^{(m)}{}_{ab} \sim |X|^{(m-n)}. 
\label{eq:extpower}
\end{eqnarray}

For brevity, we call $\mathop{h}\limits_{(n)}^{(m)}{}_{ab}$ or 
$\mathop{H}\limits_{(n)}^{(m)}{}_{ab}$ the $({}^{m}_{n})$ component and 
the matching condition for them as the $({}^{m}_{n})$ matching. 
In the matching region ($|X| \sim (GmL)^{1/2}$), the 
$({}^{m}_{n})$ component is of $O\left((Gm/L)^{(m+n)/2}\right)$. 
The matching condition requires that all the corresponding terms 
in Eqs.~(\ref{eq:bh}) and (\ref{eq:ext}) should be identical. 
Then what we have to do is to equate the terms of the same power
in $|X|$ to desired accuracy.
Thus the condition for the $({}^{m}_{n})$ matching is 
\begin{equation}
 \sum_{m'-n'=m-n \atop m'\le m} {(Gm)^{n'} \over L^{m'} } 
 \mathop{h}_{(n')}^{(m')}{}_{ab} = 
 \sum_{m'-n'=m-n \atop m'\le m} {(Gm)^{n'} \over L^{m'} }
 \mathop{H}_{(n')}^{(m')}{}_{ab}
  +O\left({(Gm)^{n+1} \over L^{m+1} }|X|^{(m-n)}\right). 
\end{equation} 

\subsection{Geodesic; $({}^0_0)$ and $({}^{1}_0)$ Matching}
We begin with the $({}^0_0)$ and $({}^{1}_0)$ matchings, 
which are of $O((Gm/L)^0)$ and of $O((Gm/L)^{1/2})$ in the 
matching region, respectively. 
First we consider the external scheme. 
In these matchings the external metric is the background itself. 
Here, the necessary order of expansion in $|X|$ is $O(|X|)$. 
Since $g_{\mu\nu}(x)dx^\mu dx^\nu =g_{\alpha\beta}(z)\bar g^\alpha{}_\mu(z,x)
\bar g^\beta{}_\nu(z,x)dx^\mu dx^\nu$ (see Eq.~(1.33) of DB), 
we get 
\begin{eqnarray}
g_{\mu\nu}(x)dx^\mu dx^\nu &=& 
\left( \left({dz \over dT}\right)^2(T) 
+2{dz^\alpha \over dT}(T){D{f}_{\alpha i} \over dT}(T)X^i
+O\left({|X|^2\over L^2}\right) \right)dT^2
\nonumber \\ && 
+2\left( {dz^\alpha \over dT}(T) {f}_{\alpha i}(T)
+{dz^\alpha \over dT}(T) {f}_{\alpha ij}(T)X^j
+{f}^{\alpha}{}_{i}(T){D{f}_{\alpha j} \over dT}(T)X^j
+O\left({|X|^2\over L^2}\right) \right)dTdX^i
\nonumber \\ && 
+\left({f}^{\alpha}_{i}(T){f}_{\alpha j}(T)
+2{f}^{\alpha}{}_{i}(T){f}_{\alpha jk}(T)X^k
+O\left({|X|^2\over L^2}\right) \right)dX^idX^j.
\label{eq:ext1}
\end{eqnarray}
Comparing the above with Eq.~(\ref{eq:ext}) and
looking at the dependence on $X$,
one can readily
extract out $\mathop{h}\limits^{(0)}_{(0)}{}_{ab}$ 
and $\mathop{h}\limits^{(1)}_{(0)}{}_{ab}$
to the lowest order in $Gm/L$.

Next we consider the internal scheme. 
The $({}^0_0)$-component is trivially given by the flat Minkowski metric. 
To know the $({}^1_0)$-component of the internal scheme, 
it is better to consider all the $({}^1_n)$-components at the same time. 
Namely we consider the linear perturbation of the black hole 
$\mathop{H}\limits^{(1)}{}_{ab}$. 
For this purpose, we consider the decomposition of the linear
perturbation in terms of the tensor harmonics as discussed in
Appendix C.
Since the timescale associated with the perturbation should be
of the order of the background curvature scale $L$, 
it is much larger than the matching radius $(GmL)^{1/2}$. 
Therefore the perturbation may be regarded as static. 
It is known that all the physical static perturbations regular on the 
black hole horizon behave as 
$\sim |X|^{J}$ asymptotically, where $J$ is the angular momentum 
eigenvalue. 
However, in $\mathop{H}\limits^{(1)}{}_{ab}$, 
there exists no term which behaves as $\sim |X|^{m}$, $(m\ge 2)$. 
Hence, except for gauge degrees of freedom,
 $\mathop{H}\limits^{(1)}{}_{ab}$ contains only $J=0$, $1$ modes.
As mentioned before, we set the perturbation 
of $J=0$, $1$ modes to vanish. Thus we conclude that we may set
$\mathop{H}\limits^{(1)}{}_{ab}=0$.

The $({}^0_0)$ matching becomes
\begin{eqnarray}
-1 &=& \left({dz \over dT}\right)^2(T) +O\left({Gm\over L}\right), 
\qquad \mbox{($TT$)-component}, \label{eq:m0tt}
\\ 
0 &=& {dz^\alpha \over dT}(T) {f}_{\alpha i}(T)+\left({Gm\over L}\right), 
\qquad \mbox{($Ti$)-components}, \label{eq:m0ti}
\\ 
\delta_{ij} &=& {f}^{\alpha}{}_{i}(T){f}_{\alpha j}(T)
+\left({Gm\over L}\right), 
\qquad \mbox{($ij$)-components}. \label{eq:m0ij}
\end{eqnarray}
Equations~(\ref{eq:m0ti}) and (\ref{eq:m0ij}) indicate 
that ${f}^{\alpha i}(T)$ are spatial triad basis along the orbit, i.e., 
\begin{equation}
 f^\alpha{}_k(T) f^\beta{}_k(T)=g^{\alpha\beta}(z(T))
 +{dz^\alpha\over dT}(T){dz^\beta\over dT}(T) +O\left({Gm\over L}\right).
\label{triad}
\end{equation}
The $({}^1_0)$ matching becomes
\begin{eqnarray}
0 &=& 
2{dz^\alpha \over dT}(T){D{f}_{\alpha i} \over dT}(T)X^i 
+\left({Gm\over L}{|X|\over L}\right), 
\qquad \mbox{($TT$)-component}, \label{eq:m1/2tt}
\\
0 &=& 
{dz^\alpha \over dT}(T) {f}_{\alpha ij}(T)X^j
+{f}^{\alpha}{}_{i}(T){D{f}_{\alpha j} \over dT}(T)X^j 
+\left({Gm\over L}{|X|\over L}\right), 
\qquad \mbox{($Ti$)-components}, \label{eq:m1/2ti}
\\ 
0 &=& 
2{f}_{\alpha (i}(T){f}^{\alpha}{}_{j)k}(T)X^k 
+\left({Gm\over L}{|X|\over L}\right), 
\qquad \mbox{($ij$)-components}. \label{eq:m1/2ij}
\end{eqnarray}

Then the covariant $T$-derivative 
of Eq.~(\ref{eq:m0tt}) and that of Eq.~(\ref{eq:m0ti}) 
with Eq.~(\ref{eq:m1/2tt}) 
result in the background geodetic motion, 
\begin{eqnarray}
{D\over dT}\left({dz^\alpha \over dT}\right)(T)=
O\left({Gm\over L}{1\over L}\right).
\label{eq:geo}
\end{eqnarray}
One can see that the internal time coordinate $T$ 
becomes a proper time of the orbit from Eq.~(\ref{eq:m0tt}) 
in the lowest order in $Gm/L$.  
In the same manner, Eq.~(\ref{eq:m1/2tt}) and 
the covariant $T$-derivative of Eq.~(\ref{eq:m0ij}) 
with $(ij)$-antisymmetric part of Eq.~(\ref{eq:m1/2ti}) give 
the geodetic transport of the triad ${f}^{\alpha}{}_{i}(T)$, 
\begin{eqnarray}
{D\over dT}{f}^{\alpha}{}_{i}(T)=O\left({Gm\over L}{1\over L}\right). 
\label{eq:para}
\end{eqnarray}
Further, from Eqs.~(\ref{eq:m1/2ti}) and (\ref{eq:m1/2ij}), 
we can see  
\begin{eqnarray}
{f}^{\alpha}{}_{ij}(T)=O\left({Gm\over L}{1\over L}\right). 
\label{eq:faij}
\end{eqnarray}

\subsection{Hypersurface Condition; $({}^2_0)$ Matching}

We now proceed to the $({}^{2}_{0})$ matching, in which the external
metric is still given by the background but there appear
non-trivial perturbations of the internal scheme.
Although it is of $O(Gm/L)$ in the matching region and 
$O((Gm/L)^{1/2})$ higher than the remaining $({}^0_1)$-component,
we consider it first for the reason which will be clarified below.

In order to obtain $\displaystyle \mathop{h}\limits^{(2)}_{(0)}{}_{ab}$,
we expand the external metric in terms of the internal coordinates up to
$O(|X|^2)$, i.e., we have to go one order higher than
Eq.~(\ref{eq:ext1}). Then the $(^2_0)$ matching becomes
\begin{eqnarray}
{1\over L^2}\mathop{H}^{(2)}_{(0)}{}_{TT}
&=& R_{\alpha\beta\gamma\delta}(z(T)){dz^\alpha\over dT}(T){f}^{\beta}{}_{i}(T)
{dz^\gamma\over dT}(T){f}^{\delta}{}_{j}(T)X^i X^j
+O\left({Gm\over L}{|X|^2\over L^2} \right),
\nonumber\\ && \mbox{($TT$)-component},
 \\ 
{1\over L^2}\mathop{H}^{(2)}_{(0)}{}_{Ti} 
&=& {1\over 2}{dz^\alpha \over dT}(T) {f}_{\alpha ijk}(T) X^j X^k
+{2\over 3}R_{\alpha\beta\gamma\delta}(z(T))
{dz^\alpha\over dT}(T){f}^{\beta}{}_{j}(T)
{f}^{\gamma}{}_{i}(T){f}^{\delta}{}_{k}(T)X^j X^k
+O\left({Gm\over L}{|X|^2\over L^2} \right),
\nonumber\\ && \mbox{($Ti$)-component},
 \\ 
{1\over L^2}\mathop{H}^{(2)}_{(0)}{}_{ij} 
&=& {f}_{\alpha (i}(T){f}^{\alpha}{}_{j)kl}(T)X^k X^l 
+{1\over 3}R_{\alpha\beta\gamma\delta}(z(T))
{f}^{\alpha}{}_{i}(T){f}^{\beta}{}_{k}(T)
{f}^{\gamma}{}_{j}(T){f}^{\delta}{}_{l}(T)X^k X^l
+O\left({Gm\over L}{|X|^2\over L^2} \right),
\nonumber\\ && \mbox{($ij$)-component},
\end{eqnarray}
where Eqs.~(\ref{eq:para}) and (\ref{eq:faij}) have been used to
simplify the expressions.
Since there appear terms which describe $J=2$ perturbations of the internal
scheme, the $({}^{2}_{0})$ matching does not determine 
$f^{\alpha}{}_{ijk}$ without specifying the gauge condition of 
the internal scheme. Here we fix it as follows.

As before, we first set $J=0$, $1$ modes of 
$\mathop{H}\limits^{(2)}_{(0)}{}_{ab}$
equal to zero. Applying the discussion given in Appendix C, we find the
$(TT)$-component does not contain $J=0$, $1$ modes. 
As for the $(Ti)$-component, it contains no $J=0$ mode but the $J=1$ mode
is present. Setting it to vanish to the lowest order in $Gm/L$, we find
\begin{equation}
{dz^\alpha\over dT}(T)f_{\alpha ikk}(T)=O\left({Gm\over L}{1\over L^3} \right).
\label{eq:fikk}
\end{equation}
Turning to the $(ij)$-component, we find the term with the Riemann
tensor contains neither $J=0$ nor $1$ mode (see Appendix C).
Hence the vanishing of $J=0$, $1$ modes implies
\begin{equation}
\left[f_{\alpha(i}(T)f^{\alpha}{}_{j)kl}(T)\right]_{J=0,1}= 
O\left({Gm\over L}{1\over L^2} \right),
\label{eq:fifjkl}
\end{equation}
where $[\cdots]_{J=0,1}$ means the $J=0$, $1$ parts of the quantity.

Let us now turn to the modes higher than $J=1$.
The $(TT)$-component contains $J=2$ mode, but it just determines the
physical perturbation of the internal metric. The higher modes are
absent from the beginning.
The $(Ti)$-component contains $J=2$ and 3 modes. As for the $J=2$ mode,
it does not involve $f^{\alpha}_{ijk}$. Hence it also gives the physical
perturbation of the internal metric. 
The $J=3$ mode of the $(Ti)$-component of 
$\mathop{h}\limits^{(2)}_{(0)}{}_{ab}$
is given by
\begin{equation}
{1\over2}{dz^\alpha\over dT}(T)f_{\alpha <ijk>}(T)X^jX^k+
 O\left({Gm\over L}{|X|^2 \over L^2} \right), 
\end{equation}
where $<ijk>$ means to take the symmetric traceless part of the indices 
(see Appendix C). 
Since this is a pure gauge degree of freedom, we may set it to vanish.
Then combining this with Eq.~(\ref{eq:fikk}), we find
\begin{equation}
{dz^\alpha\over dT}(T)f_{\alpha ijk}(T)=O\left({Gm\over L}{1\over L^2} \right).
\label{eq:fijk}
\end{equation}
We do not discuss the higher $J$ modes in the $(ij)$-component, since it
does not give us any information which is necessary 
to derive the equation of motion. 

Then, from Eqs.~(\ref{eq:m0ti}), (\ref{eq:faij}) and (\ref{eq:fijk}),
we find
\begin{equation}
 {dz^\alpha \over dT}(T)\sigma_{;\alpha}\left(x(T,X),z(T)\right)
 =-{dz^\alpha \over dT}(T){\cal F}_{\alpha}(T,X)=
 O\left({|X|^4\over L^4} L\right),
\end{equation}
to the lowest order in $Gm/L$.
Comparing this with the hypersurface condition of $\tau_x$,
Eq.~(\ref{eq:foli}), one finds that the $T=$constant hypersurface
differs from the $\tau_x=$constant hypersurface only
by $O(\epsilon^4)=O(|X|^4)$. Then following the calculations done in
section 2 again, one finds they are unaltered even if we replace 
Eq.~(\ref{eq:foli}) with
\begin{equation}
\left[\sigma_{;\alpha}(x,z(\tau))\dot z^\alpha(\tau)\right]_{\tau=\tau_x}
=O(\epsilon^4). 
\end{equation}
The only effect of this replacement is to add $O(\epsilon^4)$ to
the right hand side of the equivalent formula (\ref{eq:Afoli})
 in Appendix A.
Thus $T$ can be identified $\tau_x$ to the lowest order in $Gm/L$.
The reason why we have done the $({}^2_0)$ matching prior to the
remaining $({}^0_1)$ matching is to establish this equivalence of 
$T$ and $\tau_x$.

\subsection{External Perturbation; $({}^{0}_{1})$ Matching}
Now we proceed to the first non-trivial order in $Gm/|X|$. 
For this purpose, we must develop the external scheme. 
However, since the time slicing by the 
internal time coordinate $T$ is now identical to that by $\tau_x$ 
in the lowest order in $Gm/L$, 
we can use the previously obtained formula (\ref{eq:metper0})
for the external metric perturbation.

There remains the $({}^{0}_{1})$ matching among the  
the matchings which becomes 
of $O((Gm/L)^{1/2})$ in the matching region. 
This matching relates the masses of the particle in both schemes. 
Since this matching is of $O((|X|/L)^0)$, 
it is allowed to consider 
as if the background external metric were flat. 
As is well-known, 
the linear perturbation induced by a point-like particle 
of mass $m$ in the flat background spacetime is exactly equal to 
the asymptotic metric of a Schwarzschild black hole of mass $m$ 
in the linear order in $m$. 
This fact indicates that the matching gives a consistency condition
at this order.

In order to directly check the consistency, we rewrite 
Eq.~(\ref{eq:metper0})
in terms of the internal coordinates. From (\ref{eq:extpower}), 
we have only to consider the first term of 
the right hand side of Eq.~(\ref{eq:metper0}). 
Using Eqs.~(\ref{eq:dtrans}), (\ref{eq:465}) and
the fact that  $\epsilon=\sqrt{\sigma_{;\alpha}\sigma^{;\alpha}}
=\sqrt{{\cal F}_\alpha(T,X){\cal F}^\alpha(T,X)}$,  
the trace-reverse of Eq.~(\ref{eq:metper0}) is transformed to give 
\begin{eqnarray}
Gm\mathop{h}^{(0)}_{(1)}{}_{ab}(X)dX^a dX^b 
= Gm\left({2\over |X|}dT^2+{2\over |X|}dX^idX^i\right), 
\label{eq:10match}
\end{eqnarray}
which corresponds to the asymptotic form of the 
Schwarzschild black hole of mass $m$ 
in the harmonic coordinates. 

\subsection{Radiation Reaction; 
$({}^1_1)$ and $({}^2_1)$ Matchings}
There are many components which become of 
$O((Gm/L)^1)$ and $O((Gm/L)^{3/2})$ in the matching region.
However, we are interested in the leading order correction 
to the equation of motion with respect to $Gm/L$ and 
we found in subsection B
that in the lowest order the terms which behave as $\sim |X|^0$ 
or $|X|^1$ determines the motion of the particle. 
Therefore we consider the $({}^1_1)$ and $({}^2_1)$ matchings here. 

In order to perform the $({}^1_1)$ and $({}^2_1)$ matchings, 
the calculation in obtaining Eq.~(\ref{eq:10match}) must 
be done to the linear order in $|X|$. 
This can be done in the same way with the aid of the expression 
of $\kappa^2(x)$ given in Appendix A. 
Using further the results obtained in subsections B and C,
we find the following matching equations:
\begin{eqnarray}
{Gm\over L}\mathop{H}^{(1)}_{(1)}{}_{TT}
&=& \left\{ \left({dz \over dT}\right)^2(T)+1 \right\}
+ Gm{dz^\alpha\over dT}(T){dz^\beta\over dT}(T){\Theta}_{\alpha\beta}(T)
+O\left(\left({Gm\over L}\right)^2\right),  \label{eq:m1tt}
\\ && \mbox{($TT$)-component}, \nonumber \\ 
{Gm\over L}\mathop{H}^{(1)}_{(1)}{}_{Ti} 
&=& {dz^\alpha \over dT}(T) {f}_{\alpha i}(T)
+ Gm{dz^\alpha\over dT}(T){f}^{\beta}{}_{i}(T){\Theta}_{\alpha\beta}(T)
+O\left(\left({Gm\over L}\right)^2\right), \label{eq:m1ti}
\\ && \mbox{($Ti$)-component}, \nonumber \\ 
{Gm\over L}\mathop{H}^{(1)}_{(1)}{}_{ij} 
&=& \left\{{f}^{\alpha}{}_{i}(T){f}_{\alpha j}(T) -\delta_{ij}\right\}
+Gm{f}^{\alpha}{}_{i}(T){f}^{\beta}{}_{j}(T){\Theta}_{\alpha\beta}(T)
+O\left(\left({Gm\over L}\right)^2\right), \label{eq:m1ij}
\\ && \mbox{($ij$)-component}, \nonumber \\ 
{Gm\over L^2}\mathop{H}^{(2)}_{(1)}{}_{TT} 
&=& 2{dz^\alpha \over dT}(T){D{f}_{\alpha i} \over dT}(T)X^i 
\nonumber \\ && 
+Gm\Biggl\{
{dz^\alpha\over dT}(T){dz^\beta\over dT}(T)
f^{\gamma}{}_{i}(T){\Theta}_{\alpha\beta\gamma}(T)X^i
-{1\over 3|X|^3}f_{\alpha i}(T)f^{\alpha}{}_{jkl}(T)X^i X^j X^k X^l 
\nonumber \\ && \quad\quad
-{5\over 3|X|}R_{\alpha\beta\gamma\delta}(z(T))
{dz^\alpha\over dT}(T){f}^{\beta}{}_{i}(T)
{dz^\gamma\over dT}(T){f}^{\delta}{}_{j}(T)X^i X^j\Biggr\}
+O\left(\left({Gm\over L}\right)^2{|X|\over L}\right), \label{eq:m3/2tt} 
\\ && \mbox{($TT$)-component}, \nonumber \\ 
{Gm\over L^2}\mathop{H}^{(2)}_{(1)}{}_{Ti}
&=& {dz^\alpha \over dT}(T) {f}_{\alpha ij}(T)X^j
+{f}^{\alpha}{}_{i}(T){D{f}_{\alpha j} \over dT}(T)X^j
\nonumber \\ && 
+Gm\Biggl\{
{dz^\alpha\over dT}(T){f}^{\beta}_{i}(T)
{f}^{\gamma}_{j}(T){\Theta}_{\alpha\beta\gamma}(T)X^j
-2R_{\alpha\beta\gamma\delta}(z(T)){dz^\alpha\over dT}(T){f}^{\beta}{}_{i}(T)
{dz^\gamma\over dT}(T){f}^{\delta}{}_{j}(T)X^j 
\nonumber \\ && \quad \quad
-{2\over 3|X|}R_{\alpha\beta\gamma\delta}(z(T))
{dz^\alpha\over dT}(T){f}^{\beta}{}_{j}(T) 
{f}^{\gamma}{}_{i}(T) {f}^{\delta}{}_{k}(T)X^j X^k
\Biggr\}
+O\left(\left({Gm\over L}\right)^2{|X|\over L}\right), \label{eq:m3/2ti}
\\ && \mbox{($Ti$)-component}, 
\nonumber 
\end{eqnarray}
where
\begin{eqnarray}
 Gm {\Theta}_{\alpha\beta}(T) & := & h_{(v)\alpha\beta}(z(T)), 
\cr
 Gm {\Theta}_{\alpha\beta\gamma}(T)& := & h_{(v)\alpha\beta;\gamma}(z(T)),
\end{eqnarray}
with
\begin{equation}
 h_{(v)\mu\nu}(x):=\psi_{(v)\mu\nu}(x) 
   -{1\over 2} g_{\mu\nu}(x)\psi_{(v)}(x). 
\label{eq:hv}
\end{equation}
Note that $h_{(v)\mu\nu}(x)$ is the metric 
perturbation due to $v_{\mu\nu\alpha\beta}(x,z)$ in the Green function. 
The ($ij$)-component of the $({}^2_{1})$ matching is not presented here 
since it will not be used in the following discussion. 

According to the discussion given in subsection B we have
$\mathop{H}\limits^{(1)}_{(1)}{}_{ab}=0$. 
Thus the right hand sides of Eqs.~(\ref{eq:m1tt}), 
(\ref{eq:m1ti}) and (\ref{eq:m1ij}) must vanish. 
As for $\mathop{H}\limits^{(2)}_{(1)}{}_{ab}$, 
we cannot set them equal to zero. 
However, again from the condition that 
there are no $J=0$, $1$ modes, the corresponding parts 
in the right hand sides of Eqs.~(\ref{eq:m3/2tt}) and (\ref{eq:m3/2ti})
must vanish if they are extracted out. 
Following the discussion given in Appendix C,
one finds the terms involving the Riemann tensor contains no $J=0,1$ mode.
Furthermore, the $J=0,1$ mode of the term involving $f^{\alpha}{}_{jkl}$ in
Eq.~(\ref{eq:m3/2tt}) is proportional to
\begin{equation}
\left[f_{\alpha(i}(T)f^{\alpha}{}_{j)kl}(T)\right]_{J=0,1}\,,
\end{equation}
which vanishes at the lowest order in $Gm/L$ by Eq.~(\ref{eq:fifjkl}).
Hence the $J=1$ modes of the remaining terms in Eqs.~(\ref{eq:m3/2tt}) and
(\ref{eq:m3/2ti}) are extracted out to give
\begin{eqnarray} 
0 &=& 
2{dz^\alpha \over dT}(T){D{f}_{\alpha i} \over dT}(T) 
+ Gm\,{dz^\alpha\over dT}(T){dz^\beta\over dT}(T)
{f}^{\gamma}{}_{i}(T){\Theta}_{\alpha\beta\gamma}(T)
+O\left(\left({Gm\over L}\right)^2{1\over L}\right),
\label{eq:m3/2ttg}
\\ && \mbox{($TT$)-component}, \nonumber \\ 
0 &=& 
{f}_{\alpha [i}(T){D{f}^{\alpha}{}_{j]} \over dT}(T)
+ Gm\,{\Theta}_{\alpha\beta\gamma}(T){dz^\alpha\over dT}(T)
 {f}^{\beta}{}_{[i}(T){f}^{\gamma}{}_{j]}(T)
+O\left(\left({Gm\over L}\right)^2{1\over L}\right),
\label{eq:m3/2tig}
\\ && \mbox{($Ti$)-component}.
 \nonumber 
\end{eqnarray}
The $J=0$ mode of Eq.~(\ref{eq:m3/2tt}) is shown to vanish,
while that of Eq.~(\ref{eq:m3/2ti}) exists but 
it does not contain useful information. 
The matching condition for this mode just gives 
the equation which determines $(dz^{\alpha}/dT)f_{\alpha ii}$
to the first order in $Gm/L$. 

The covariant $T$-derivative of Eq.~(\ref{eq:m1tt}) and 
that of Eq.~(\ref{eq:m1ti}) with Eq.~(\ref{eq:m3/2ttg}) 
give the equation of motion with leading correction 
due to the radiation reaction, 
\begin{eqnarray}
{D\over dT}{dz^\alpha\over dT}(T) = 
-{Gm\over 2}
\left({\Theta}^\alpha{}_{\beta\gamma}(T)
+{\Theta}^\alpha{}_{\gamma\beta}(T)-{\Theta}_{\beta\gamma}{}^\alpha(T)\right)
{dz^\beta\over dT}(T){dz^\gamma\over dT}(T)
+O\left(\left({Gm\over L}\right)^2{1\over L}\right). 
\label{eq:damp1}
\end{eqnarray}
The leading correction to the evolution of the `triad' basis,
 $f^{\alpha}{}_{i}(T)$, are also obtained from Eq.~(\ref{eq:m3/2ttg}) 
and the covariant $T$-derivatives of Eq.~(\ref{eq:m1ij}) 
with Eq.~(\ref{eq:m3/2tig}) as  
\begin{eqnarray}
{D\over dT}{f}^{\alpha}{}_{i}(T)=
- {Gm\over 2}
\left({\Theta}^\alpha{}_{\beta\gamma}(T)
+{\Theta}^\alpha{}_{\gamma\beta}(T)-{\Theta}_{\beta\gamma}{}^\alpha(T)\right)
{f}^{\beta}{}_{i}(T){dz^\gamma\over dT}(T)
 +O\left(\left({Gm\over L}\right)^2{1\over L}\right).
\label{eq:damp2}
\end{eqnarray}

Since the internal time coordinate $T$ is not properly normalized 
in the external metric, 
we define the proper time, $\tau=\tau(T)$, such that $(dz/d\tau)^2 =-1$. 
It is easy to see that we should choose 
\begin{equation}
 {d\tau\over dT}=1+{Gm\over 2}\Theta_{\alpha\beta}(\tau)
 {dz^{\alpha}\over d\tau}(\tau){dz^{\beta}\over d\tau}(\tau)
 +O\left(\left({Gm\over L}\right)^2\right).
\end{equation}
Since the second term on the right hand side of this equation 
is proportional to the small perturbation induced by the particle, it is
 guaranteed to stay small even after a long time interval 
compared with the reaction time scale $\tau_r=O\left((Gm/L)^{-1}L\right)$. 
Then Eq.~(\ref{eq:damp1}) becomes 
\begin{equation}
{D\over d\tau}{dz^\alpha\over d\tau}(\tau) 
= -{Gm\over 2}\left({dz^\alpha\over d\tau}{dz^\beta\over d\tau}
{dz^\gamma\over d\tau}{dz^\delta\over d\tau}
+2g^{\alpha\beta}(z){dz^\gamma\over d\tau}{dz^\delta\over d\tau}
-g^{\alpha\delta}(z){dz^\beta\over d\tau}{dz^\gamma\over d\tau}
\right)(\tau)~{\Theta}_{\beta\gamma\delta }(\tau)
+O\left(\left({Gm\over L}\right)^2{1\over L}\right).
\label{eq:damp1r}
\end{equation}
It is easy to see that this equation is identical to 
that obtained at the end of the previous section.

Also, the triad basis are not properly normalized in the external metric. 
Thus we define, $e^{\alpha}{}_{i}(\tau)$, 
such that $e_{\alpha i}(\tau)e^{\alpha}{}_{j}(\tau)=\delta_{ij}$ and   
$e^{\alpha}{}_{i}(\tau)=(\delta_{ij}+s_{ij})f^{\alpha}{}_{j}
-Gm({dz^{\alpha} / dT})({dz^{\beta} / dT})f^\gamma{}_i
{\Theta}_{\beta\gamma}$, where 
$s_{ij}$ is of $O(Gm/L)$ and the last term is added 
so as to satisfy the orthonormal condition, 
$e_{\alpha i}(\tau)({dz^{\alpha} / d\tau})(\tau)=0$ 
(see Eq.~(\ref{eq:m1ti})). We find 
\begin{equation}
 s_{ij}=- {Gm\over 2}\Theta_{\alpha\beta}(\tau)
 {f^{\alpha}{}_i}(\tau){f^{\beta}{}_j}(\tau)
 +O\left(\left({Gm\over L}\right)^2\right),
\end{equation}
and again this is guaranteed to stay small.
Then the evolution of the normalized triad $e^{\alpha}{}_{i}(\tau)$ becomes 
\begin{equation}
{D \over d\tau}e^{\alpha}{}_{i}(\tau) 
= - {Gm\over 2} \left({dz^\alpha\over d\tau}{dz^\beta\over d\tau}
e^{\gamma}{}_{i}{dz^\delta\over d\tau} 
+g^{\alpha\beta}(z){dz^\gamma\over d\tau}e^{\delta}{}_{i} 
-g^{\alpha\delta}(z) e^{\beta}{}_{i}{dz^\gamma\over d\tau} 
\right) (\tau)~ {\Theta}_{\beta\gamma\delta }(\tau) 
+O\left(\left({Gm\over L}\right)^2{1\over L}\right).
\label{eq:damp2r}
\end{equation}

\section{Implications}

In this section, we first 
consider the physical meaning of the 
equation of motion obtained in the preceding two sections. 
Since the equation of motion we have obtained contains
an unknown function $v_{\mu\nu\alpha\beta}(x,z)$, 
we need to give a method to explicitly 
determine the particle trajectory. 
Here we consider a couple of possibilities to calculate it 
in the case of a specific background, such as Kerr spacetime. 

In order to make the meaning of the equation 
of motion manifest, 
we divide the perturbed metric in the external scheme  
into two pieces as 
\begin{equation}
 h_{\mu\nu}(x)=h_{(u)\mu\nu}(x)+h_{(v)\mu\nu}(x), 
\end{equation}
where $h_{(v)\mu\nu}(x)$ is the part due to the tail term (defined in 
Eq.~(\ref{eq:hv})) and $h_{(u)\mu\nu}(x)$ is the part
due to the $u_{\mu\nu\alpha\beta}$ term in 
the Green function (corresponding
to the first term on the right hand side of Eq.~(\ref{eq:metper}). 
The singular behavior of the perturbed metric 
in the coincidence limit is totally due to $h_{(u)\mu\nu}(x)$. 
Thus, we introduce the regularized 
perturbed spacetime defined by 
\begin{eqnarray}
\tilde g_{(v)\mu\nu}(x):=g_{\mu\nu}(x)+h_{(v)\mu\nu}(x),
\label{eq:regmet}
\end{eqnarray}
which has no singular behavior any more. 
Then we find the equation of motion (\ref{eq:damp1}) and the 
evolution equation of the triad basis (\ref{eq:damp2}) 
coincide with the geodesic equation and the geodetic parallel transport,
respectively, in this regularized perturbed spacetime, 
$\tilde g_{(v)\mu\nu}$. To see this let us consider
 the parallel transport of a vector 
$A^{\alpha}$ along a geodesic $x^\alpha=z^\alpha(\tilde\tau)$
in this spacetime. It is given by
\begin{equation}
 {\tilde D_{(v)}\over d\tilde\tau}A^{\alpha}:=
  {D\over d\tilde\tau}A^{\alpha}+\delta 
  \Gamma_{(v)}{}^{\alpha}{}_{\beta\gamma} A^{\beta}
  {dz^{\gamma}\over d\tilde\tau}=0, 
\label{eq:saigo}
\end{equation}
to the linear order in $h_{(v)\nu\mu}$ where 
\begin{equation}
  \delta \Gamma_{(v)}{}^{\alpha}{}_{\beta\gamma}:=
   {1\over 2}\left(h_{(v)}{}^{\alpha}{}_{\beta;\gamma}+
   h_{(v)}{}^{\alpha}{}_{\gamma;\beta}-h_{(v)\beta\gamma}{}^{;\alpha}
   \right). 
\end{equation}
Then 
one recovers Eqs.~(\ref{eq:damp1}) and (\ref{eq:damp2}) by 
identifying $\tilde\tau$ with $T$ and 
replacing $A^\alpha$ with $dz^\alpha/d T$ or $f^{\alpha}{}_{i}$.
This fact is the main result of this paper. 

We note that the present equation of motion is analogous to that in 
the electromagnetic case, 
except that the instantaneous 
reaction force which is proportional to higher derivatives of 
the particle velocity is absent in the present case. 
This is because the particle traces the 
geodesic in the lowest order of approximation. 
If an external force field exists, the assumption of the 
geodesic motion in the lowest order breaks down and furthermore 
the contribution of the external force field 
to the energy momentum tensor must be taken into account.
Since this fact makes the problem too complicated,
it is beyond the scope of the present paper. 

Now let us consider the way how to construct $\tilde g_{(v)\mu\nu}$. 
As stated in the beginning of this section, 
in order to evaluate the particle trajectory explicitly,
a practical scheme to calculate $\tilde g_{(v)\mu\nu}$ 
must be developed. 
Unfortunately, we do not have any scheme 
which can be satisfactorily applied when calculating the particle 
trajectory with the effect of the radiation reaction, even 
on a specific background spacetime such as a Kerr black hole. 
Here we just give a few primitive discussions on this matter. 
For definiteness, we focus on the case of the retarded boundary
condition.

Basically, there seems to be two approaches
for calculating $\tilde g_{(v)\mu\nu}$ (or equivalently 
$h_{(v)\mu\nu}$). 
The first one is to calculate $h_{(v)\mu\nu}$ directly. 
The second one is to calculate 
$h_{\mu\nu}=h_{(u)\mu\nu}+h_{(v)\mu\nu}$ and subtract 
$h_{(u)\mu\nu}$ from it. 
In the following, we discuss only the first approach. 
As for the second approach, we have  
nothing to mention here, but
this direction of research may be fruitful. 

By definition, $h_{(u)\mu\nu}$ evaluated on the particle trajectory 
is independent of the past history of the particle.\footnote{
There is a possibility that 
the future light cone emanating from $z$ crosses 
the particle trajectory. Since inclusion of this possibility
 makes the problem too complicated, we do not consider it here.}
Therefore if we consider the metric defined by 
\begin{equation}
 h^{(\Delta\tau)}_{\mu\nu}(x)=
  \left(\delta_{\mu}{}^{\rho}\delta_{\nu}{}^{\sigma}
    -{1\over 2}g_{\mu\nu}(x) g^{\rho\sigma}(x)\right)
   \int_{-\infty}^{\tau_x-\Delta\tau}~
  d\tau' G^{Ret}_{\rho\sigma\alpha\beta}(x,z(\tau'))\dot z^{\alpha}
   (\tau')\dot z^{\beta}(\tau'), 
\end{equation}
for any finite $\Delta\tau$ $(>0)$, it will not contain 
$h_{(u)\mu\nu}$ when it is evaluated on the particle trajectory. 
The difference between $h^{(\Delta\tau)}_{\mu\nu}$ and 
$h_{(v)\mu\nu}$ comes from the integral 
over the small interval, 
\begin{equation}
   \sim Gm \int_{\tau_x-\Delta\tau}^{\tau_x}~
  d\tau' v_{\rho\sigma\alpha\beta}(x,z(\tau'))\dot z^{\alpha}
   (\tau')\dot z^{\beta}(\tau').
\end{equation}
Since $v_{\rho\sigma\alpha\beta}(x,z)$ is a regular function, 
this integral will be negligible for a sufficiently small 
$\Delta\tau$. 
Thus $\lim_{\Delta\tau\to0}h^{(\Delta\tau)}_{\mu\nu}$ will give 
$h_{(v)\mu\nu}$.

In the case of the electromagnetic (vector) Green function,
a calculation along the above strategy was performed by
DeWitt and DeWitt\cite{DeW2} by assuming
the background gravitational field is weak so that
its metric is given by the small perturbation 
on the Minkowski metric,
\begin{equation}
 g_{\mu\nu}=\eta_{\mu\nu}+h^{(b)}_{\mu\nu}\,.
\end{equation}
DeWitt and DeWitt calculated the relevant part of the
Green function perturbatively to the first order in $h^{(b)}_{\mu\nu}$
by using the Minkowski Green function.
An analogous calculation seems possible in the case of
gravity to evaluate $h_{(v)\mu\nu}$, 
though it seems difficult to develop such calculations
to higher orders in $h^{(b)}_{\mu\nu}$.

As an alternative approach, we give below
a speculation of a new method in which we do not 
have to assume the deviation from the 
Minkowski spacetime to be small.
The standard perturbation formalism of a Kerr black hole 
gives us the retarded Green function schematically in 
the form,\footnote{Note that this expression is only schematic, 
just for the present explanation. The correct expression can be 
fond in \cite{Gal}.}
\begin{equation}
 G^{Ret}_{\mu\nu\alpha\beta}(x,z)\sim
  \sum_{j,m}\int d\omega {1\over W_{j,m,\omega}}
   e^{-i\omega(t_x-t_z)} 
   h^{j,m,\omega}_{\mu\nu}(r_x,\theta_x,\varphi_x) 
   h^{j,m,\omega}_{\alpha\beta}(r_z,\theta_z,\varphi_z), 
\end{equation}
where $h^{j,m,\omega}_{\mu\nu}(r,\theta,\varphi)$ 
represents the spatial mode functions with appropriate boundary 
conditions and $W_{j,m,\omega}$ is the Wronskian.
Here we stress the following point. 
Provided that $x$ is inside the future light 
cone emanating from $z$, 
the integrand becomes exponentially small as 
the imaginary part of $\omega$ goes to $-\infty$. 
Thus the contour of the $\omega$ integration 
can be closed in the lower half complex
plane of $\omega$. 
Since the integrand has poles at the quasi-normal mode frequencies, 
at which $W_{j,m,\omega}$ vanishes, the $\omega$-integral 
can be transformed into the summation over the 
contributions from the residues of the poles.
We denote this expression of the retarded Green function by
\begin{equation}
 G^{Ret(QN)}_{\mu\nu\alpha\beta}(x,z) 
 \sim
  -2\pi i\sum_{j,m,n} 
   \left({dW_{j,m,\omega}\over d\omega}\right)^{-1}_{\omega=\omega_n}
   e^{-i\omega_n (t_x-t_z)} 
   h^{j,m,\omega_n}_{\mu\nu}(r_x,\theta_x,\varphi_x) 
   h^{j,m,\omega_n}_{\alpha\beta}(r_z,\theta_z,\varphi_z), 
\end{equation}
where $\omega_n$ are the quasi-normal mode frequencies. 
On the other hand, when $x$ coincides with $z$, 
the replacement of the $\omega$-integration 
into the summation over the quasi-normal modes is not allowed.
Therefore $\displaystyle
\lim_{x\rightarrow z} G^{Ret(QN)}_{\mu\nu\alpha\beta}(x,z)$
is not equal to $\displaystyle
\lim_{x\rightarrow z} G^{Ret}_{\mu\nu\alpha\beta}(x,z)$.
In particular, we expect that $G^{Ret(QN)}_{\mu\nu\alpha\beta}(x,z)$ 
is regular in the coincidence limit $x\to z$, 
since $(dW_{j,m,\omega}/d\omega)_{\omega=\omega_n}$ 
is known to behave as $1/n!$\cite{ShuWi}. Thus it may 
give the regularized retarded Green function which we want to obtain.

Before closing this section, we give comments on 
some proposals to the equation of motion including 
the effect of the gravitational radiation reaction.
One is the use of the radiative Green function 
(a half of the difference between the retarded and 
advanced Green functions) 
in the case of a Kerr background\cite{Gal}
As seen easily by using the results obtained in the 
previous section, the use of the radiative Green function 
instead of the retarded one results 
in the replacement of $\psi_{(v)\mu\nu}(x)$
by $\psi^{Rad}_{(v)\mu\nu}(x)$, which is defined by 
\begin{equation}
\psi^{Rad}_{(v)}{}_{\beta\gamma}(x):=-Gm\int^{+\infty}_{-\infty}d\tau' 
v_{\beta\gamma\alpha'\beta'}(x,z(\tau'))
\dot z^{\alpha'}(\tau')\dot z^{\beta'}(\tau'). 
\end{equation}
Gal'tsov has proved that 
the back reaction force computed using the radiative Green function 
correctly gives the loss of 
the energy and the z-component of the 
angular momentum of the particle 
in quasi-periodic orbits. 
However, we do not think that this fact indicates 
the correctness of the prescription 
because those constants of motion are special ones 
which reflect the existence of the corresponding 
Killing vector field. 
For example, it is still uncertain if the 
radiative Green function is useful in evaluating 
radiation reaction effect on the Carter constant. 
We would like to come back to this point in future 
publication. 

Recently Ori has argued also in the case of a Kerr background 
that the partial wave decomposition 
of the retarded Green function with respect 
to the spheroidal harmonics as schematically shown above 
would result in 
the automatic removal of the divergence
when averaged over several orbital periods\cite{Ori},
thus making it possible to derive the radiation reaction 
to the Carter constant.
We do not think his argument works. 
Let us consider expansion of the Newtonian 
potential of a particle off the origin of the coordinates 
by using the spherical harmonics. 
If one calculate the self-force on the particle, one finds the 
contribution from each partial wave is finite. 
However, the net self-force is intrinsically ill-defined and 
if one sums up all the contributions from 
the different modes, it is easy to see that thus 
obtained self force diverges.
Thus the finiteness of the partial wave contributions
does not imply that of their sum.

\section{Conclusion}
In this paper, we have derived the equation of motion of a particle 
on a given background
including the effect of the gravitational radiation reaction,
i.e., with corrections of order $(Gm)$ where $m$ is the mass of 
the particle.
Although we use the terminology `radiation reaction' here, 
it may not be appropriate in a strict sense 
because the equation of motion we have derived
may well contain something more than just the radiation reaction. 
In fact, in the electromagnetic case, the existence of 
the effect which can be termed as
`the induced polarization force on the background spacetime'
is reported by many authors\cite{inpol}. The existence and the physical
meaning of it in the gravitational case is left for future study.

We have derived the equation of motion in two different ways. 
First we have considered an extension of the electromagnetic 
counter part developed by DeWitt and Brehme\cite{DeWitt}. 
Due to the nature of gravitational interaction,
there appears an ambiguity concerning the renormalization of mass of the
particle, which cannot be resolved within this approach. We have then
derived the equation of motion by setting an ansatz which seems
physically reasonable without justification. 
In order to overcome this problem, we have developed a method based on
 the matched asymptotic expansion, assuming the local geometry around the
particle is described by a spherically symmetric black hole plus tidal
perturbations. This latter approach has proved to 
be very powerful and we have succeeded in obtaining the same equation of
motion as obtained in the former approach.

The correction term of $O(Gm)$ is 
found to be entirely given by the part of the metric perturbation
which is due to the tail term of the Green function.
Defining the regularized metric as 
the background plus this tail part of the perturbation,
we have found that the equation of motion is the geodesic equation
on this regularized perturbed metric.

It is important to note that the particle does not have to be a black
hole but the resultant equation of motion can be equally applicable 
to any compact bodies such as neutron stars.
The essential assumption here is that the only scale associated with
the particle is $Gm$ and the structure is basically spherically
symmetric. In this sense, we have shown the strong equivalence principle to
the first order in $Gm$.

We also emphasize that our result gives a justification of
 the so-called black hole perturbation approach for the first time.
In the black hole perturbation approach, one calculates the
gravitational radiation from a particle orbiting a black hole with the
assumption that the particle is a point-like object with the energy
momentum tensor described by the delta function. Although this approach
has been fruitful, there has been always skepticism about the validity
of the delta functional source. What we have shown in this paper justifies
the use of the delta function in the source energy momentum tensor.

In this paper, we has considered a particle which is essentially
structureless. A next step will be to include the intrinsic angular
momentum of the particle, i.e, to consider a spinning particle. 
As we have found the method of the matched asymptotic expansion is very
powerful, there is a big possibility that we can extend the present
analysis to the case of a spinning particle. Extension in this direction
is under study.

\acknowledgments
We thank T. Futamase, T. Nakamura, M. Shibata,  H. Tagoshi, 
K.S. Thorne and A.G. Wiseman for fruitful conversations. 
Y. M. thanks Prof. H. Sato and Prof. S. Ikeuchi for their 
continuous encouragements. This work was supported in part 
by Monbusho Grant-in-Aid for Scientific Research No. 5427.

\appendix

\section{Notation and basic formulas} 
Here in this appendix, we summarize our notation and basic 
formulas. As for the formulas already derived in DB, we just 
write down the results.
\subsection{\bf Basic Notation}
\noindent
(i) The Riemann tensor and the Ricci tensor are defined by 
\begin{eqnarray}
R^\mu{}_{\nu\xi\rho}&:=&\Gamma^\mu{}_{\nu\rho,\xi}-\Gamma^\mu{}_{\nu\xi,\rho}
+\Gamma^\mu{}_{\sigma\xi}\Gamma^{\sigma}{}_{\nu\rho}-\Gamma^\mu{}_{\sigma\rho}
 \Gamma^{\sigma}{}_{\nu\xi}, 
\\
R_{\mu\nu}&:=&R^\xi{}_{\mu\xi\nu}, 
\\
R&:=&R^\mu{}_{\mu}. 
\end{eqnarray}

\noindent
(ii) Symmetrization and anti-symmetrization of the tensor indices are 
described by  
$t^{(\mu\nu)}=(t^{\mu\nu}+t^{\nu\mu})/2$ and 
$t^{[\mu\nu]}=(t^{\mu\nu}-t^{\nu\mu})/2$, respectively. 

\noindent
(iii) For an arbitrary bi-tensor $Q$,
\begin{eqnarray}
Q_{;\alpha}(x,z(\tau_x)) & := & [Q_{;\alpha}(x,z)]_{z=z(\tau_x)}, 
\cr
Q_{;\mu}(x,z(\tau_x)) & := & [Q_{;\mu}(x,z)]_{z=z(\tau_x)}, 
\end{eqnarray}
while 
\begin{equation}
[Q(x,z(\tau_x))]_{;\mu} := Q_{;\mu}(x,z(\tau_x)) 
  +Q_{;\alpha}(x,z(\tau_x)) \dot z^{\alpha}(\tau_x) \tau_{x;\mu}. 
\end{equation}

\noindent
(iv) The basic equations satisfied by a half the squared 
geodetic interval $\sigma (x,z)$ (Eqs.~(1.11) and (1.12) of DB):
\begin{eqnarray}
&&\sigma(x,z)={1\over 2} g^{\mu\nu}(x)\sigma_{;\mu}(x,z)\sigma_{;\nu}(x,z)=
{1\over 2} g^{\alpha\beta}(z)\sigma_{;\alpha}(x,z)\sigma_{;\beta}(x,z),
\nonumber\\
&& \lim_{x\rightarrow z}\sigma (x,z)=0.
\label{111db}
\end{eqnarray}

\noindent
(v) The defining equations of the parallel displacement bi-vector
(Eqs.~(1.31) and (1.32) of DB):
\begin{eqnarray}
&&\bar g_{\mu\alpha;\nu}(x,z)g^{\nu\sigma}(x) \sigma_{;\sigma}(x,z) =0,
\quad
\bar g_{\mu\alpha;\beta}(x,z)g^{\beta\gamma}(z) 
\sigma_{;\gamma}(x,z) =0,
\nonumber\\
&&\lim_{x\rightarrow z}\bar g_{\mu}{}^{\alpha} (x,z) 
= \delta_{\mu}{}^{\alpha}.
\label{131db}
\end{eqnarray}

\noindent
(vi) The definition of $\Delta(x,z)$ 
(Eqs.~(1.50), (1.51), (1.60) and (1.61) of DB):
\begin{equation}
\Delta(x,z):=|\bar g^{\alpha\mu}(z,x)\sigma_{;\mu\beta}(x,z)|. 
\label{Deltadef}
\end{equation}

\subsection{Basic formulas}

\noindent
(i) (Eq.~(1.28) of DB)
\begin{equation}
 \sigma_{;\alpha\beta}(x,z)=g_{\alpha\beta}(z)+ {1\over 3} 
 R_{\alpha}{}^{\gamma}{}_{\beta}{}^{\delta}(z) 
 \sigma_{;\gamma}(x,z) \sigma_{;\delta}(x,z) +O(\epsilon^3).
\label{128db}
\end{equation}

\noindent
(ii)
\begin{eqnarray}
 \bar g^{\mu\alpha}{}_{;\beta}(x,z) & = & {1\over 2} 
 \bar g^{\mu\gamma}(x,z)R^{\alpha}{}_{\gamma\beta\delta}(z) 
 \sigma^{;\delta}(x,z) +O(\epsilon^2), \cr
 \bar g^{\mu\alpha}{}_{;\nu}(x,z) & = & {1\over 2} 
 \bar g^{\mu\beta}(x,z) \bar g_{\nu}{}^{\gamma}(x,z) 
 R^{\alpha}{}_{\beta\gamma\delta}(z) 
 \sigma^{;\delta}(x,z) +O(\epsilon^2), 
\label{140db}
\end{eqnarray}
which can be obtained from Eqs.~(1.40) and (1.41) of DB. 

\noindent
(iii) (Eqs.~(1.51), (1.64) and (1.73) of DB)
\begin{equation}
 \sigma_{;\mu\beta}(x,z)=-\bar g_{\mu}{}^{\alpha}(x,z)
 \left(g_{\alpha\beta}(z)-{1\over 6} 
 R_{\alpha\gamma\beta\delta}(z) 
 \sigma^{;\gamma}(x,z) \sigma^{;\delta}(x,z)\right)+O(\epsilon^3).
\label{173db}
\end{equation}

\noindent
(iv) Here we give the formulas which we need to expand 
Eq.~(\ref{eq:metper}) to obtain Eq.~(\ref{eq:metper0}).
Since we want to express Eq.~(\ref{eq:metper}) 
in terms of $\tau_x$ defined by Eq.~(\ref{eq:foli}) instead 
of $\tau_{Ret/Adv}(x)$, 
we expand each factor of each term which consists of Eq.~(\ref{eq:metper}) 
with $\delta_{Ret/Adv}(x)$, (\ref{eq:retard}). 
We first consider 
$\left[\dot\sigma(x,z(\tau))\right]_{\tau=\tau_{Ret/Adv}(x)}$, 
which is expanded as 
\begin{eqnarray}
\left[\dot\sigma(x,z(\tau))\right]_{\tau=\tau_{Ret/Adv}(x)}
&=&\dot\sigma(x,z(\tau_x))+\ddot\sigma(x,z(\tau_x))\delta_{Ret/Adv}(x)
\nonumber \\ && \qquad 
+{1\over 2}\stackrel{...}{\sigma}(x,z(\tau_x))\delta_{Ret/Adv}^2(x)
+{1\over 3!}\stackrel{....}{\sigma}(x,z(\tau_x))\delta_{Ret/Adv}^3(x)
+O(\epsilon^4). 
\end{eqnarray}
Each term is computed as follows; 
\begin{eqnarray}
\dot\sigma(x,z(\tau_x))
&=&\sigma_{;\alpha}(x,z(\tau_x))\dot z^{\alpha}(\tau_x)=0,
\label{eq:Afoli}\\
\ddot\sigma(x,z(\tau_x))\,&=:&-\kappa^2(x)
\nonumber\\
&=&\sigma_{;\alpha\beta}(x,z(\tau_x))\dot z^\alpha(\tau_x)\dot z^\beta(\tau_x)
+\sigma_{;\alpha}(x,z(\tau_x))\ddot z^{\alpha}(\tau_x)
\nonumber \\ 
& = & 
\left(g_{\alpha\beta}(z(\tau_x))+{1\over 3}
 R_{\alpha}{}^{\gamma}{}_{\beta}{}^{\delta}(z(\tau_x)) 
 \sigma_{;\gamma}(x,z(\tau_x)) \sigma_{;\delta}(x,z(\tau_x))\right)
  \dot z^{\alpha}(\tau_x)\dot z^{\beta}(\tau_x)
\nonumber \\ 
&& \quad +
 \sigma_{;\alpha}(x,z(\tau_x)) \ddot z^{\alpha}(\tau_x)
 +O(\epsilon^3),
\nonumber \\ 
\\ 
\stackrel{...}{\sigma}(x,z(\tau_x))
&=&\sigma_{;\alpha}(x,z(\tau_x))\stackrel{...}{z}{}^\alpha(\tau_x)
+O(\epsilon^2),
\\ 
\stackrel{....}{\sigma}(x,z(\tau_x))
&=& -g_{\alpha\beta}(z(\tau_x))
\ddot z^\alpha(\tau_x)\ddot z^\beta(\tau_x)+O(\epsilon). 
\end{eqnarray}
In the above computation, 
we used Eqs.~(\ref{128db}), (\ref{eq:foli}), 
and the normalization condition 
$(dz/d\tau)^2 = -1+ O(Gm/L)$, 
which was proved to be consistent in sections 3 and 4. 
Then we obtain 
\begin{equation}
\left[{1\over\dot\sigma(x,z(\tau))}\right]_{\tau=\tau_{Ret/Adv}(x)}
=\pm{1\over\epsilon(x)\kappa(x)}\biggl(1
\mp{1\over 3}\epsilon(x)\stackrel{...}{z}{}^\alpha(\tau_x)
\sigma_{;\alpha}(x,z(\tau_x))
-{1\over 8}\epsilon^2(x)\ddot z^2(\tau_x)
+O(\epsilon^3)\biggr).
\end{equation} 
Noting the explicit form of $u^{\mu\nu\,\alpha\beta}(x,z)$ 
in Eq.~(\ref{eq:u}), it is necessary to compute 
$[\Delta^{1/2}(x,z(\tau))]_{\tau=\tau_{Ret/Adv}(x)}$, 
$[\bar g_{\mu\alpha}(x,z(\tau))]_{\tau=\tau_{Ret/Adv}(x)}$ and 
$[\dot z^\alpha(\tau)]_{\tau=\tau_{Ret/Adv}(x)}$. 
In the same way, 
\begin{eqnarray}
&& \biggl[\Delta^{1/2}(x,z(\tau))\biggr]_{\tau=\tau_{Ret/Adv}(x)}
=1+O(\epsilon^3),
\\
&& \biggl[\bar g_{\mu\alpha}(x,z(\tau))\biggr]_{\tau=\tau_{Ret/Adv}(x)}
\nonumber \\ && \qquad 
=\bar g_{\mu\alpha}(x,z(\tau_x))
\pm{1\over 2}\bar g_\mu{}^\beta(x,z(\tau_x))
R_{\alpha\beta\gamma\delta}(z(\tau_x))
\sigma^{;\gamma}(x,z(\tau_x))\dot z^\delta(\tau_x)\epsilon(x)
+O(\epsilon^3), 
\\
&& \biggl[\dot z^\alpha(\tau)\biggr]_{\tau=\tau_{Ret/Adv}(x)} 
=\dot z^\alpha(\tau_x)\mp\epsilon(x)\kappa^{-1}(x)\ddot z^\alpha(\tau_x) 
+{1\over 2}\epsilon^2(x)\stackrel{...}{z}{}^\alpha(\tau_x)+O(\epsilon^3), 
\end{eqnarray}
Putting them into the first term in the parentheses of 
Eq.~(\ref{eq:metper}), 
and using Eq.~(\ref{eq:ulim}) and the fact 
\begin{equation}
 A^\mu(x)=\bar g^\mu{}_\alpha(x,z)
 \biggl(A^\alpha(z)-\sigma_{;\beta}(x,z)A^{\alpha;\beta}(z)
+O(\epsilon^2)\biggr),
\end{equation} 
in computing the second term, 
we obtain Eq.~(\ref{eq:metper0}). 


\section{2nd order variation of the Einstein tensor}
We derive Eq.~(\ref{eq:2ndein}) 
by taking the variation of the Einstein-Hilbert action. 
We first compute the Einstein-Hilbert action 
of the metric, $\tilde g_{\mu\nu}=g_{\mu\nu}+h_{\mu\nu}$. 
For this purpose we define a differential operator $\delta_{{\bf g}}$: 
\begin{eqnarray}
&& \delta_{{\bf g}}Q(g)
=\lim_{\epsilon\to0}{Q(g+\epsilon h)-Q(g)\over\epsilon},
\\ 
&& \delta^2_{{\bf g}}Q(g)
=\lim_{\epsilon\to0}{Q(g+\epsilon h)+Q(g-\epsilon h)-2Q(g)\over2\epsilon^2}\,.
\end{eqnarray}
We first note the 1st variation of the Einstein tensor 
taken from a standard text book,
\begin{eqnarray}
G^{(1)\mu\nu}
&=&(-h^{\mu\nu;\xi}+h^{\xi\mu; \nu}+h^{\xi\nu; \mu})_{;\xi}
-h^{;\mu\nu}
-g^{\mu\nu}(h^{\xi\rho}{}_{;\xi\rho}-h^{;\xi}{}_{;\xi})
\\ 
&=&(-\psi^{\mu\nu;\xi}+\psi^{\xi\mu; \nu}
+\psi^{\xi\nu; \mu})_{;\xi}
-g^{\mu\nu}\psi^{\xi\rho}{}_{;\xi\rho}\,,
\label{eq:ein1var}
\end{eqnarray}
where $\psi_{\mu\nu}=h_{\mu\nu}-(1/2)g_{\mu\nu}h$.

Since we are interested only in the 2nd order variation, 
we compute the terms proportional to ${\bf h}^2$ in the action:
\begin{eqnarray}
\delta^2_{{\bf g}}(\sqrt{-g}R)
&=&(\delta^2_{{\bf g}}\sqrt{-g})R
+(\delta_{{\bf g}}\sqrt{-g}\delta_{{\bf g}}g^{\mu\nu}+
+\sqrt{-g}\delta^2_{{\bf g}}g^{\mu\nu})R_{\mu\nu}
\nonumber\\ &&
+(\sqrt{-g}\delta_{{\bf g}}g^{\mu\nu}
+g^{\mu\nu}\delta_{{\bf g}}\sqrt{-g}) 
\delta_{{\bf g}}R_{\mu\nu}
+\sqrt{-g}g^{\mu\nu}\delta^2_{{\bf g}}R_{\mu\nu}
\nonumber\\
&=&-\sqrt{-g}(h^{\mu\nu}-{1\over2}g^{\mu\nu}h)
\delta_{{\bf g}}R_{\mu\nu}
+\sqrt{-g}g^{\mu\nu}\delta^2_{{\bf g}}R_{\mu\nu},
\label{2ndvar}
\end{eqnarray}
where we have used the assumption that the background spacetime is
vacuum, $R_{\mu\nu}=0$.
Now
\begin{eqnarray}
\delta_{{\bf g}}R_{\mu\nu}
&=&(\delta_{{\bf g}}\Gamma^\xi{}_{\mu\nu})_{;\xi}
-(\delta_{{\bf g}}\Gamma^\xi{}_{\mu\xi})_{;\nu}\,,
\\
\delta^2_{{\bf g}}R_{\mu\nu}
&=&(\delta^2_{{\bf g}}\Gamma^\xi{}_{\mu\nu})_{;\xi}
-(\delta^2_{{\bf g}}\Gamma^\xi{}_{\mu\xi})_{;\nu}
+\delta_{{\bf g}}\Gamma^\xi{}_{\rho\xi}
\delta_{{\bf g}}\Gamma^\rho{}_{\mu\nu}
-\delta_{{\bf g}}\Gamma^\xi{}_{\rho\nu}
\delta_{{\bf g}}\Gamma^\rho{}_{\mu\xi}\,.
\end{eqnarray}
Inserting these into Eq.~(\ref{2ndvar}) and using
\begin{eqnarray}
\delta_{{\bf g}}\Gamma^\xi{}_{\mu\nu}={1\over2}
(h^\xi{}_{\mu; \nu}+h^\xi{}_{\nu; \mu}-h_{\mu\nu}{}^{;\xi}),
\end{eqnarray}
we obtain the 2nd order variation of the Einstein-Hilbert action,
\begin{eqnarray}
L^{(2)}
&=&{1\over 16\pi G}{1\over\sqrt{-g}}\delta^2_{{\bf g}}(\sqrt{-g}R)
\nonumber\\
&&\quad={1\over 64\pi G}[
-h_{\mu\nu;\xi}h^{\mu\nu;\xi}+2h_{\mu\nu;\xi}h^{\xi\mu; \nu}
-2h_{\mu\nu}{}^{;\nu}h^{;\mu}+h_{;\mu}h^{;\mu}]
\nonumber\\
&&\quad={1\over 64\pi G}[
-\psi_{\mu\nu;\xi}\psi^{\mu\nu;\xi}
+2\psi_{\mu\nu;\xi}\psi^{\xi\mu; \nu}
+{1\over2}\psi_{;\mu}\psi^{;\mu}],
\end{eqnarray}
where $\psi=\psi^\alpha{}_\alpha=-h$ and
we have discarded unimportant total divergence terms.

We note 
\begin{eqnarray}
{\delta S\over\delta\tilde g_{\mu\nu}}[\tilde {\bf g}]
={\delta S\over\delta g_{\mu\nu}}[{\bf g}+\delta {\bf g}].
\end{eqnarray}
Thus the 2nd order variation of the Einstein tensor can be obtained 
by taking the variation of the action with respect to $g_{\mu\nu}$:
\begin{eqnarray}
{\delta\over\delta g_{\mu\nu}}\int L^{(2)}\sqrt{-g}d^4x
={1\over 16\pi G}\int d^4x\sqrt{-g}\left(-G^{(2)\mu\nu}\right)
\end{eqnarray}
Hence,
\begin{equation}
G^{(2)\mu\nu}
=-8\pi G\left[2{\delta L^{(2)}\over \delta g_{\mu\nu}}
+g^{\mu\nu}L^{(2)}\right].
\end{equation}
Carrying out the variation of $L^{(2)}$ one finds

\begin{eqnarray}
G^{(2)\mu\nu}
=&&-{1\over2}\biggl[{1\over2}
(\psi^{\mu\xi;\rho}+\psi^{\mu\rho;\xi}-\psi^{\xi\rho;\mu})
(\psi^{\nu}{}_{\xi;\rho}+\psi^{\nu}{}_{\rho;\xi}
-\psi_{\xi\rho}{}^{;\nu})
\nonumber\\
&&\qquad+\psi^{\mu\xi}{}_{;\rho\xi}\psi^{\rho\nu}
+\psi^{\nu\xi}{}_{;\rho\xi}\psi^{\rho\mu}
-\psi^{\mu\nu}{}_{;\xi\rho}\psi^{\xi\rho}
-(\psi^{\mu\nu}\psi^{\xi\rho}{}_{;\rho})_{;\xi}
-{1\over4}\psi^{;\mu}\psi^{;\nu}
\nonumber\\
&&\qquad+{1\over4}g^{\mu\nu}
[-\psi_{\rho\sigma;\xi}\psi^{\rho\sigma;\xi}
+2\psi_{\rho\sigma;\xi}\psi^{\xi\rho;\sigma}
+{1\over2}\psi_{;\xi}\psi^{;\xi}]\biggr]
\nonumber\\
&&\qquad+{1\over2}\psi\, G^{(1)\mu\nu}.
\end{eqnarray}
Note that if $\bf h$ is a linear perturbation which satisfies
$G^{(1)\mu\nu}[{\bf h}]=0$, the tensor
$T^{\mu\nu}_{G}:=-(1/8\pi G)G^{(2)\mu\nu}$ describes the conserved
energy momentum tensor of the perturbed gravitational field.

\section{Tensor harmonics expansion}
Here we briefly review the construction of the scalar 
and the vector harmonics in terms of the 
symmetric trace-free (STF) tensor \cite{BlaDam}. 
We introduce the notation 
\begin{equation}
 A_{<i_1 i_2\cdots i_\ell>}, 
\end{equation}
to represent the totally symmetric and trace-free 
part of $A_{i_1 i_2\cdots i_\ell}$.
More explicitly in the cases of $\ell=2$, $3$, 
\begin{eqnarray}
 A_{<ij>} & = & A_{(ij)}-{1\over 3}\delta_{ij} A_{kk}, \cr
 A_{<ijk>} & = & A_{(ijk)}-{1\over 5}\left(
 \delta_{ij} A_{(kmm)}+\delta_{jk} A_{(imm)}+
 \delta_{ki} A_{(jmm)}\right).
\end{eqnarray}

The spherical harmonics expansion of a scalar function 
$A$ on the unit-sphere can be written as 
\begin{equation}
 A=\sum_{\ell=0}^{\infty} A_{<i_1 i_2\cdots i_\ell>}
 n^{<i_1}n^{i_2}\cdots n^{i_\ell>}, 
\end{equation}
where $n^i=X^i/|X|$. 
In this case, the order $\ell$, which is associated with the angular
dependence, is equivalent to the total angular momentum, $J$. 
Thus the $J$ mode of the $(TT)$-component 
of the metric perturbation is totally determined by 
its angular dependence. 
Namely, the terms in the $(TT)$-component 
of the metric perturbation which contain
\begin{equation}
 1,\quad n^i,\quad n^{<i}n^{j>}, 
\end{equation}
correspond to the $J=0$, $1$, $2$ modes, respectively. 

Next we consider the expansion of a vector field $A_i$, 
\begin{equation}
 A_i=\sum_{\ell=0}^{\infty} A_{i <i_1 i_2\cdots i_\ell>}
 n^{<i_1}n^{i_2}\cdots n^{i_\ell>}. 
\end{equation}
In this case the term of the $\ell$-th order in the angular dependence
is decomposed into $J=\ell+1$, $\ell$ and $\ell-1$.
This is done by using the Clebsch-Gordan reduction formula \cite{BlaDam}, 
\begin{equation}
 U_i T_{i_1 i_2\cdots i_\ell}= R^{(+)}_{i<i_1 i_2\cdots i_\ell>}
  +{\ell\over \ell+1} \epsilon_{ji<i_\ell} 
   R^{(0)}_{i_1 i_2\cdots i_{\ell-1}>j}
  +{2\ell-1\over 2\ell+1} \delta_{i<i_\ell} 
   R^{(-)}_{i_1 i_2\cdots i_{\ell-1}>}, 
\end{equation}
where $T_{i_1 i_2\cdots i_\ell}$ is a STF tensor of order $\ell$ and 
\begin{eqnarray}
    R^{(+)}_{i_1 i_2\cdots i_{\ell+1}} & := & 
    U_{<i_{\ell+1}} T_{i_1 i_2\cdots i_\ell>},
\cr
    R^{(0)}_{i_1 i_2\cdots i_{\ell}} & := & 
     U_{j} T_{k<i_1 i_2\cdots i_{\ell-1}} \epsilon_{i_\ell>jk}, 
\cr 
    R^{(-)}_{i_1 i_2\cdots i_{\ell-1}} & := & 
     U_{j} T_{j i_1 i_2\cdots i_{\ell-1}}. 
\end{eqnarray}
We perform the decomposition explicitly for $\ell\le 2$ here. 
For $\ell=0$, there exists no $J=0$ mode and 
it trivially corresponds to the $J=1$ mode. 
For $\ell=1$, the decomposition is performed as 
\begin{equation}
 A_{ij}n^j= \left[\left(A_{(ij)}
  -{1\over 3}\delta_{ij} A_{kk}\right) + A_{[ij]} 
  +{1\over 3}\delta_{ij} A_{kk}\right] n^j, 
\end{equation}
and the first, second and third terms in the square brackets 
correspond to the $J=2$ ,$1$ and $0$ modes, respectively. 
{}For $\ell=2$, we obtain the decomposition formula as 
\begin{equation}
 A_{i<jk>}n^{<j} n^{k>}= \left[A_{<ijk>}+{2\over 3}
  \epsilon_{mi<j} B^{(2)}_{k>m} + {3\over 5} 
  \delta_{i<j} B^{(1)}_{k>}\right]
  n^{<j} n^{k>}, 
\label{c10}
\end{equation}
where
\begin{eqnarray}
 B^{(2)}_{ij} & = & {1\over 2}
 (A_{k<mi>}\epsilon_{jkm}+A_{k<mj>}\epsilon_{ikm}),
\cr 
 B^{(1)}_{k} & = & A_{i<jk>}\delta_{ij}\,,
\label{c9}
\end{eqnarray}
and the first, second and third terms correspond to the $J=3$, $2$
and $1$ modes, respectively.

As an example, let us consider a vector, 
\begin{equation}
 {dz^{\alpha}\over dT} R_{\alpha\beta\gamma\delta}
 f^{\beta}{}_{j} f^{\gamma}{}_{i} f^{\delta}{}_{k} X^j X^k,
\end{equation}
which appears in the $({}^2_0)$ and $({}^2_1)$ matchings in section 4.
First we decompose it in terms of its angular dependence as 
\begin{equation}
 {dz^{\alpha}\over dT} R_{\alpha\beta\gamma\delta}
 f^{\gamma}{}_{i} 
 \left( f^{\beta}{}_{<j} f^{\delta}{}_{k>} X^{<j} X^{k>} 
 +{1\over 3} f^{\beta}{}_{k} f^{\delta}{}_{k} |X|^{2}\right). 
\label{c12}
\end{equation}
Using the relation 
(\ref{triad}) 
and the fact that the Ricci tensor vanishes,
the second term in the parentheses is rewritten as 
\begin{equation}
 {1\over 3} {dz^{\alpha}\over dT} R_{\alpha\beta\gamma\delta}
 f^{\gamma}{}_{i} 
 {dz^{\beta}\over dT} {dz^{\delta}\over dT} |X|^{2}, 
\end{equation}
and is found to be zero due to 
the symmetry of the Riemann tensor.  
Thus we have only to consider the first term 
in the parenthesis of Eq.~(\ref{c12}), which is 
decomposed further with the aid of the formulas
(\ref{c9}) and (\ref{c10}) as
\begin{equation}
 {dz^{\alpha}\over dT} R_{\alpha\beta\gamma\delta}
 \left(f^{\gamma}{}_{<i} 
 f^{\beta}{}_{j} f^{\delta}{}_{k>} +
 {2\over 3}\epsilon_{mi<j} 
 F^{(2)\gamma\beta\delta}_{k>m} +
 {3\over 5} \delta_{i<j} F^{(1)\gamma\beta\delta}_{k>}
 \right)
 X^{<j} X^{k>},
\end{equation}
where
\begin{eqnarray}
F^{(2)\gamma\beta\delta}_{ij} & := & 
 {1\over 2}\left( f^{\gamma}{}_{m} f^{\beta}{}_{<n} 
  f^{\delta}{}_{i>} \epsilon_{jmn}
+ f^{\gamma}{}_{m} f^{\beta}{}_{<n} 
  f^{\delta}{}_{j>} \epsilon_{imn}\right), 
\cr
F^{(1)\gamma\beta\delta}_{i} & := & 
 {1\over 2}\left( f^{\gamma}{}_{k} f^{\beta}{}_{i} 
  f^{\delta}{}_{k} + f^{\gamma}{}_{k} f^{\beta}{}_{k} 
  f^{\delta}{}_{i} \right) -{1\over 3} 
  f^{\gamma}{}_{i} f^{\beta}{}_{k} f^{\delta}{}_{k}. 
\end{eqnarray}
It is easy to see that the first and third terms 
vanish due to the symmetry of the Riemann tensor and 
the Ricci flatness. 
Thus only the $J=2$ mode remains. 

As for a tensor field, it is not necessary for us to give
general discussion here.
The only term which requires our consideration is 
\begin{equation}
  R_{\alpha\beta\gamma\delta}
 f^{\alpha}{}_{i} f^{\beta}{}_{k} 
 f^{\gamma}{}_{j} f^{\delta}{}_{m} X^k X^m\,,
\end{equation}
which appears in the $(ij)$-component of the $({}^2_0)$ matching.
In this case, it is better to use the symmetry of 
the Riemann tensor from the beginning. 
First, we define the spatial triad components of
 the Riemann tensor by 
\begin{equation}
 R_{ijkm}:=R_{\alpha\beta\gamma\delta}
 f^{\alpha}{}_{i} f^{\beta}{}_{j} 
 f^{\gamma}{}_{k} f^{\delta}{}_{m}\,.
\end{equation}
Introducing a symmetric tensor defined by 
\begin{equation}
 {\cal R}_{ij}={1\over 4} \epsilon_{ikm} \epsilon_{jns} 
      R_{kmns}, 
\end{equation} 
we can express $R^{ikjm}$ in terms of ${\cal R}_{ij}$ as
\begin{equation}
R_{ijkm} = \epsilon^{nij} \epsilon^{skm} {\cal R}_{ns}. 
\end{equation}
Then the symmetric tensor 
${\cal R}_{ij}$ is decomposed into STF tensors as 
\begin{equation}
{\cal R}_{ij}={\cal R}_{<ij>}+ 
              {1\over 3}\delta_{ij}{\cal R}_{kk}. 
\label{calRab}
\end{equation}
Counting the number of indices, we find that 
the first and second terms in Eq.~(\ref{calRab}) 
correspond to $J=2$ and $0$ modes, respectively. 
However, again owing to the symmetry of the Riemann tensor and 
the Ricci flatness, the $J=0$ mode vanishes 
and only the $J=2$ mode remains. 



\begin{thebibliography}{ederf}
\bibitem{Thorne} K. S. Thorne, in {\it Proceedings of the 8th 
Nishinomiya-Yukawa Meomorial Symposium: Relativistic Cosmology}, ed. 
M. Sasaki (Universal Academy Press, Tokyo, 1994), p67, and the 
references therein.

\bibitem{LISA} K. Danzmann et al., 
{\it `LISA: Proposal for a Laser-Interferometric Gravitational Wave Detector 
in Space'}, MPQ-177 (1993), unpublished. 

\bibitem{Last} C. Cutler et al., Phys. Rev. Lett. {\bf 70}, 
2984 (1994), and the references therein. 

\bibitem{bhp} H. Tagoshi, M. Shibata, T. Tanaka and M. Sasaki, 
Phys. Rev. {\bf D} {\it in press}, 
Caltech preprint GRP-434/Osaka University preprint OU-TAP 28, 
gr-qc/9603028, and the references therein. 


\bibitem{DeWitt} B. S. DeWitt and R. W. Brehme, Ann. Phys. 
(N.Y.) {\bf 9}, 220 (1960).

\bibitem{Death} P. D. D'Eath, Phys. Rev. {\bf D11}, 1387 (1975).

\bibitem{Thorne1} K. S. Thorne, and J. B. Hartle, Phys. Rev. {\bf D31},
 1815 (1985).

\bibitem{Zerilli} F. J. Zerilli, Phys. Rev. {\bf D2}, 2141 (1970).

\bibitem{DeW2} C. M. DeWitt and B. S. DeWitt, Physics {\bf 1}, 3 (1964). 

\bibitem{ShuWi} B. F. Shutz and C. M. Will, Astrophys. J. {\bf 291}, 
L33 (1985). 

\bibitem{Gal} D. V. Gal'tsov, J. Phys. {\bf A15}, 3737 (1982).

\bibitem{Ori} A. Ori, Phys. Lett. A {\bf 202}, 347 (1995).

\bibitem{inpol} M. Carmeli, Phys. Rev. {\bf 138}, B1003 (1965);\\
 A. G. Smith and C. M. Will, Phys. Rev. {\bf D22}, 1276 (1980).

\bibitem{BlaDam} e.g., Appendix A of L. Blanchet and T. Damour, 
Philos. Trans. R. Soc. London {\bf A320}, 379 (1986). 
















\end{thebibliography}
\end{document}